\documentclass[12pt,a4paper]{article}
\pdfoutput=1
\usepackage{color}
\usepackage{amssymb,amsmath,bm,bbm}
\usepackage{epsf}
\usepackage{epsfig}
\usepackage{afterpage}
\usepackage{longtable}
\usepackage[dvipsnames]{xcolor}
\usepackage[linktoc=page,bookmarks=false,colorlinks=false,
linkbordercolor=RoyalBlue,citebordercolor=ForestGreen,urlbordercolor=CornflowerBlue]{hyperref}
\usepackage{latexsym,mathrsfs,dsfont}
\usepackage[normalem]{ulem} 
\usepackage[compress]{cite}
\usepackage{graphicx}
\usepackage{url}
\usepackage{paralist}
\usepackage{bbold}

\newcommand{\mvs}{\vbox{\vskip 8mm}}

\newcommand{\ord}{\mathcal{O}}

\newcommand{\IM}{{\rm Im}}
\newcommand{\RE}{{\rm Re}}

\newcommand{\gev}{\, {\rm GeV}}
\newcommand{\mev}{\, {\rm MeV}}

\newcommand{\bsi}{B_6^{(1/2)}}
\newcommand{\bei}{B_8^{(3/2)}}

\newcommand{\phm}{\phantom{--}}
\def\epe{\varepsilon'/\varepsilon}
\newcommand{\beq}{\begin{equation}}
\newcommand{\eeq}{\end{equation}}
\newcommand{\be}{\begin{equation}}
\newcommand{\ee}{\end{equation}}
\newcommand{\bi}{\begin{itemize}}
\newcommand{\ei}{\end{itemize}}
\newcommand{\ba}{\begin{array}}
\newcommand{\ea}{\end{array}}
\newcommand{\beqa}{\begin{eqnarray}}
\newcommand{\eeqa}{\end{eqnarray}}
\newcommand{\bea}{\begin{eqnarray}}
\newcommand{\eea}{\end{eqnarray}}
\newcommand{\beqn}{\begin{eqnarray}}
\newcommand{\eeqn}{\end{eqnarray}}
\newcommand{\al}{\alpha}

\definecolor{red}{cmyk}{0,1,1,0.4}

\def\kpn{K^+\rightarrow\pi^+\nu\bar\nu}
\def\klpn{K_{L}\rightarrow\pi^0\nu\bar\nu}

\usepackage{fancyhdr}
\advance \headheight by 3.0truept       

\begin{document}


\vspace{-14mm}
\begin{flushright}
        {FLAVOUR(267104)-ERC-100 \\ LTH 1051}
\end{flushright}

\vspace{8mm}

\begin{center}
{\Large\bf
\boldmath{Improved anatomy of $\epe$ in the Standard Model}}
\\[12mm]
{\bf Andrzej~J.~Buras,${}^1$ Martin Gorbahn,${}^2$ Sebastian J{\"a}ger${}^3$
and Matthias Jamin${}^4$ \\[1cm]}
{\small
${}^1$TUM Institute for Advanced Study, Lichtenbergstr.~2a, D-85748 Garching, Germany\\
Physik Department, TU M\"unchen, James-Franck-Stra{\ss}e, D-85748 Garching, Germany\\[2mm]
${}^2$Department of Mathematical Sciences, University of Liverpool, Liverpool, L69 7ZL, UK\\[2mm]
${}^3$Department of Physics and Astronomy, University of Sussex, Brighton, BN1 9QH, UK\\[2mm]
${}^4$Instituci\'o Catalana de Recerca i Estudis Avan\c cats (ICREA), IFAE,\\
Universitat Aut\`onoma de Barcelona, E-08193 Bellaterra (Barcelona), Spain}
\end{center}

\vspace{6mm}

\begin{abstract}
\noindent
We present a new analysis of the ratio $\varepsilon'/\varepsilon$ within the
Standard Model (SM) using a formalism that is manifestly independent of the
values of leading $(V-A)\otimes(V-A)$ QCD penguin, and EW penguin hadronic
matrix elements of the operators $Q_4$, $Q_9$, and $Q_{10}$, and applies to
the SM as well as extensions with the same operator structure. It is valid
under the assumption that the SM exactly describes the data on CP-conserving
$K\to \pi\pi$ amplitudes. As a result of this and the high precision now
available for CKM and quark mass parameters, to high accuracy
$\varepsilon'/\varepsilon$ depends only on two non-perturbative parameters,
$B_6^{(1/2)}$ and $B_8^{(3/2)}$, and perturbatively calculable Wilson
coefficients. Within the SM, we are separately able to determine the hadronic
matrix element $\langle Q_4 \rangle_0$ from CP-conserving data, significantly
more precisely than presently possible with lattice QCD. Employing
$B_6^{(1/2)}=0.57\pm 0.19$ and $B_8^{(3/2)}=0.76\pm 0.05$, extracted from
recent results by the RBC-UKQCD collaboration, we obtain
$\varepsilon'/\varepsilon=(1.9\pm 4.5)\times 10^{-4}$, substantially more
precise than the recent RBC-UKQCD prediction and $2.9\,\sigma$ below
the experimental value $(16.6\pm 2.3)\times 10^{-4}$, with the error being fully
dominated by that on $B_6^{(1/2)}$. Even discarding lattice input completely,
but employing the recently obtained bound $B_6^{(1/2)}\le B_8^{(3/2)}\le 1$
from the large-$N$ approach, the SM value is found more than $2\,\sigma$ below
the experimental value. At $B_6^{(1/2)}=B_8^{(3/2)}=1$, varying all other
parameters within one sigma, we find
$\varepsilon'/\varepsilon=(8.6\pm 3.2)\times 10^{-4}$. We present a detailed
anatomy of the various SM uncertainties, including all sub-leading hadronic
matrix elements, briefly commenting on the possibility of underestimated SM
contributions as well as on the impact of our results on new physics models.
\end{abstract}

\setcounter{page}{0}
\thispagestyle{empty}
\newpage

\tableofcontents

\newpage

\section{Introduction}

One of the important actors of the 1990s in particle physics was the ratio
$\epe$  that measures the size of the direct CP violation in $K_L\to\pi\pi$ 
relative to the indirect CP violation described by $\varepsilon_K$. In the
Standard Model (SM), $\varepsilon^\prime$ is governed by QCD penguins but 
receives also an important destructively interfering  contribution from
electroweak penguins that is generally much more sensitive to new physics (NP)
than the QCD penguin contribution. Reviews on $\epe$ can be found in 
\cite{Bertolini:1998vd,Buras:2003zz,Pich:2004ee,Cirigliano:2011ny,Bertolini:2012pu}.

A long-standing challenge in  making predictions for $\epe$ within the SM and
its extensions has been the strong interplay of QCD penguin contributions
and electroweak penguin contributions to this ratio. In the SM, QCD penguins
give a positive contribution and electroweak penguins a negative one. In order
to obtain a useful prediction for $\epe$, the relevant contributions of the QCD
penguin and electroweak penguin operators must be know accurately.

As far as short-distance contributions (Wilson coefficients of QCD and 
electroweak penguin operators) are concerned, they have been known already
for more than twenty years at the NLO level
\cite{Buras:1991jm,Buras:1992tc,Buras:1992zv,Ciuchini:1992tj,Buras:1993dy,Ciuchini:1993vr}
and present technology could extend them to the NNLO level if necessary. First
steps in this direction have been taken in \cite{Buras:1999st,Gorbahn:2004my,Brod:2010mj}.

The situation with hadronic matrix elements is another story and even if 
significant progress on their evaluation has been made  over the last 25 years,
the present status is clearly not satisfactory as we will discuss below. But,
already in 1993, an approach has been proposed in  \cite{Buras:1993dy} which,
as far as $\epe$ is concerned, avoids direct calculation of some of the most
difficult hadronic matrix elements. It assumes that the real parts of the
isospin amplitudes $A_0$ and $A_2$, which exhibit the $\Delta I=1/2$ rule,
are fully described by SM dynamics and their experimental values are used to
determine to a very good approximation hadronic matrix elements of all
$(V-A)\otimes (V-A)$ operators, among them the so-called $Q_4$ QCD penguin
operator. While not as important as the $(V-A)\otimes (V+A)$ QCD penguin and
electroweak penguin operators, $Q_6$ and $Q_8$, the operator $Q_4$ has been
known since the early days of analyses of $\epe$ \cite{Buchalla:1989we,Ciuchini:1992tj,Buras:1993dy}
to be responsible for a significant part of the suppression of this ratio. In
the presence of a partial cancellation of the positive contribution of $Q_6$ to
$\epe$ by the one of $Q_8$, an accurate determination of the contribution from
$Q_4$ and from  the electroweak penguin operators $Q_9$ and $Q_{10}$  to $\epe$
by means of CP-conserving data was an important virtue of our approach. 

Another virtue of our approach is based on the fact that in the SM the
amplitudes ${\rm Re} A_0$ and ${\rm Re} A_2$ originate already at tree-level.
Similar to the observables used for tree-level determination of CKM parameters,
also relevant for $\epe$, they are expected to be only marginally affected by NP
contributions. Whether NP could contribute to ${\rm Re} A_0$ and ${\rm Re} A_2$
at some level is an interesting question, to which we will return briefly in
Section~\ref{sec:explBSM}. But, for the time being we assume that they are
fully dominated by SM dynamics.

With the contribution of $(V-A)\otimes (V-A)$ operators being determined from
the data on ${\rm Re} A_0$ and $ {\rm Re} A_2$ it was possible to write down
an analytic formula for $\epe$ that incorporated all NLO QCD and QED corrections
and summarised the remaining dominant hadronic uncertainty in terms of two
parameters $\bsi$ and $\bei$ that parametrise the relevant matrix elements
of the dominant operators $Q_6$ and $Q_8$ and have to be calculated using a
non-perturbative framework like lattice QCD or the large-$N$ approach
\cite{Bardeen:1986uz,Buras:2014maa}. They cannot be extracted from
CP-conserving data as their contributions to ${\rm Re} A_0$ and $ {\rm Re} A_2$
are marginal at $\mu\approx m_c$ used in the approach of \cite{Buras:1993dy}.
In fact one of the reasons for choosing the value $\mu=m_c$ was to eliminate
them from the determination of the matrix elements of $(V-A)\otimes (V-A)$ 
operators {from the CP-conserving data.}

Over the last twenty years the basic formula for $\epe$ of \cite{Buras:1993dy}
has been improved \cite{Buras:1996dq,Bosch:1999wr,Buras:2003zz} due to the
increased accuracy in the value of the QCD coupling and other input parameters,
like the values of $m_t$ and $m_s$. We refer to
\cite{Buras:1996dq,Bosch:1999wr,Buras:2003zz}, where useful information on our
approach can be found. The most recent version of our analytic formula has
been presented in \cite{Buras:2014sba,Buras:2015qea}.

One new aspect of the present paper is the realisation that under the
assumption that NP contributions to ${\rm Re} A_0$ and ${\rm Re} A_2$ are
negligible, the leading contributions of $(V-A)\otimes (V-A)$ operators to
$\epe$ can be entirely expressed in terms of their Wilson coefficients.
Furthermore, we derive a formula for $\epe$ which under the above assumption
can be used in any extension of the SM in which the operator structure is the
same as in the SM. NP enters only through the modified values of the Wilson
coefficients and the dominant non-perturbative uncertainties are contained in 
\be\label{nonperturbative}
\bsi,\qquad \bei, \qquad
q \equiv \frac{z_+(\mu)\langle Q_+(\mu) \rangle_0}{z_-(\mu)\langle Q_-(\mu)
\rangle_0} \,.
\ee
The ratio $q$, involving matrix elements of current-current operators $Q_{\pm}$
and their Wilson coefficients $z_\pm$, enters the determination of the
contribution of  $(V-A)\otimes (V-A)$ operators from CP-conserving data and
its range will be estimated in Section~\ref{sec:2}. But for $0\le q\le 0.1$
obtained from  QCD lattice and large-$N$ approaches the dependence of $\epe$
on $q$ is very weak.

As far as the parameters $\bsi$ and $\bei$ are concerned, $\bsi=\bei=1$ in the
large-$N$ limit of QCD. The study of $1/N$ corrections to the large-$N$ limit
indicated that $\bei$ is suppressed below unity \cite{Hambye:1998sma}, but no
clear-cut conclusion has been reached in that paper on $\bsi$. Moreover, the
precise amount of suppression  of $\bei$ could not be calculated in this
approach. Fortunately, in the meantime significant progress has been achieved
in the case of the matrix element $\langle Q_8\rangle_2$  by the RBC-UKQCD
lattice collaboration \cite{Blum:2015ywa}, which allowed to determine $\bei$
to be \cite{Buras:2015qea}
\be\label{Lbei}
B_8^{(3/2)}(m_c)=0.76\pm 0.05\, \qquad (\mbox{RBC-UKQCD}),
\ee
in agreement with large-$N$ expectations \cite{Hambye:1998sma,Buras:2015xba}, but with
higher precision. 

But also some progress on $\bsi$ has been made, both by lattice QCD and the
large-$N$ approach. In particular, very recently the RBC-UKQCD lattice
collaboration \cite{Bai:2015nea} presented their first result for the matrix
element $\langle Q_6\rangle_0$ from which one can extract (see below and
\cite{Buras:2015xba})
\be\label{Lbsi}
B_6^{(1/2)}(m_c)=0.57\pm 0.19\, \qquad (\mbox{RBC-UKQCD}).
\ee
This low value of $\bsi$ is at first sight surprising and as it is based on a
numerical simulation one could wonder whether it is the result of a statistical
fluctuation. But the very recent analysis in the large-$N$ approach in
\cite{Buras:2015xba} gives strong support to the values in (\ref{Lbei}) and (\ref{Lbsi}).
In fact, in this analytic approach one can demonstrate explicitly the
suppression of both $\bsi$ and $\bei$ below their large-$N$ limit $\bsi=\bei=1$
and  derive a conservative upper bound on both $\bsi$ and $\bei$ which reads
\cite{Buras:2015xba}
\be\label{NBOUND}
\bsi\le \bei < 1 \, \qquad (\mbox{\rm large-}N).
\ee
While one finds $B_8^{(3/2)}(m_c)=0.80\pm 0.10$, the result for $\bsi$ is less
precise but there is a strong indication that $\bsi < \bei$ in agreement with
(\ref{Lbei}) and (\ref{Lbsi}). For further details, see \cite{Buras:2015xba} and
Section~\ref{sec:2a} below.

Employing the lattice results of (\ref{Lbei}) and (\ref{Lbsi}), in our
numerical analysis we find 
\be\label{LBGJJ}
   \epe = (1.9 \pm 4.5) \times 10^{-4} \,,
\ee
consistent with, but significantly more precise than the result obtained
recently by the RBC-UKQCD lattice collaboration \cite{Bai:2015nea},
\be\label{RBC}
(\epe)_\text{SM} = (1.4 \pm 7.0)\times 10^{-4} \,.
\ee
This is even more noteworthy considering the fact that our result comprises
also uncertainties from isospin corrections and CKM parameters which were not
considered in the error estimate of \cite{Bai:2015nea}. Our result differs
with close to $3\,\sigma$ significance from the experimental world average
from NA48 \cite{Batley:2002gn} and KTeV
\cite{AlaviHarati:2002ye,Worcester:2009qt} collaborations, 
\be\label{EXP}
(\epe)_\text{exp}=(16.6\pm 2.3)\times 10^{-4} \,,
\ee
suggesting evidence for new physics in $K$ decays.

But even discarding the lattice results, varying all input parameters, we find
at the bound $\bsi=\bei=1$,
\be\label{BoundBGJJ}
(\epe)_\text{SM}= (8.6\pm 3.2) \times 10^{-4} \,,
\ee
still $2\,\sigma$  below the experimental data. We consider this bound
conservative since employing the lattice value in (\ref{Lbei}) and
$\bsi=\bei=0.76$, instead of (\ref{BoundBGJJ}), one obtains
$(6.0\pm 2.4)\times 10^{-4}$. 

This already shows that with the rather precise value of $\bei$ from 
lattice QCD, the final result for $\epe$ dominantly depends on the value of
$\bsi$ and both  lattice QCD \cite{Bai:2015nea} and the large-$N$ approach
\cite{Buras:2015xba} indicate that the SM value of $\epe$ is significantly below the
data.

The two main goals of the present paper are:
\begin{itemize}
\item
Derivation  of a new version of our
formula for $\epe$ which could also be used beyond the SM and which appears
to be more useful than its variants presented by us in the past.
\item
Demonstration that our approach provides a substantially more accurate
prediction for $\epe$ in the SM than it is presently possible within lattice
QCD and that the upper bound in (\ref{BoundBGJJ}) is rather conservative.
\end{itemize}

It should be stressed that assuming dominance of SM dynamics in CP-conserving
data, our determination of the contributions of  $(V-A)\otimes (V-A)$ operators
to $\epe$ is basically independent of the non-perturbative approach used.
The RBC-UKQCD lattice collaboration calculates these contributions directly
and we will indeed identify a significant difference between their estimate
of the $Q_4$ contribution to $\epe$ and ours. 

Our paper is organised as follows. In Section~\ref{sec:2}, we derive the
analytic formula for $\epe$ in question using the strategy of
\cite{Buras:1993dy} but improving on it. Using this formula, we present a new
analysis of $\epe$ within the SM exhibiting its sensitivity to the precise
value of $\bsi$ and the weak dependence on $q$. In Section~\ref{sec:2a}, we
perform the anatomy of uncertainties affecting $\epe$ and present the
prediction of $\epe$ in the SM, including a discussion of its $\bsi$
dependence. In Section~\ref{sec:3}, we extract from the lattice-QCD results
of \cite{Bai:2015nea} the values of the most important hadronic matrix elements
and compare them with ours. This allows us to identify the main origin of the
difference between (\ref{LBGJJ}) and (\ref{RBC}). In particular, we point out
an approximate correlation between the contribution of the $Q_4$ operator to
$\epe$ and the value of $\RE A_0$ valid in any non-perturbative approach. In
Section~\ref{sec:explSM}, we investigate if thus far neglected SM contributions
could bring our result for $\epe$ into agreement with the experimental findings.
A brief general discussion of the impact of possible NP contributions to
${\rm Re} A_{0,2}$ and $ {\rm Im} A_{0,2}$ and
of the implications of our results for NP models is given in
Section~\ref{sec:explBSM}. The summary of our observations and an outlook are
presented in Section~\ref{sec:5}. In Appendix~\ref{app:subleading}, we discuss
the sub-leading contributions to our prediction for $\epe$ and in
Appendix~\ref{app:X}, for completeness, an updated analytic formula for $\epe$
in the SM is presented in the form used in several of our papers in the past
(e.g. \cite{Buras:2015qea}) that is equivalent to the one derived in
Section~\ref{sec:2}, but exhibits the $m_t$, $\alpha_s$, $m_s$ and $m_d$
dependences more explicitly.


\section{Basic formulae}\label{sec:2}

\subsection{Effective Hamiltonian}

We use the effective Hamiltonian for $\Delta S=1$ transitions of
\cite{Buras:1991jm,Buras:1992tc,Buras:1992zv,Ciuchini:1992tj,Buras:1993dy,Ciuchini:1993vr}
\begin{equation}
{\cal H}_\text{eff} = \frac{G_F}{\sqrt{2}}\, V_{ud} V_{us}^* \sum_{i=1}^{10}
\big( z_i(\mu) + \tau \; y_i(\mu) \big)\, Q_i(\mu) \,,\quad
\tau \equiv -\,\frac{V_{td}V_{ts}^*}{V_{ud}V_{us}^*} \,.
\label{eq:Heff}
\end{equation}

The contributing operators are given as follows:

{\bf Current--Current:}
\begin{equation}\label{O1} 
Q_1 = (\bar s_{\alpha} u_{\beta})_{V-A}\;(\bar u_{\beta} d_{\alpha})_{V-A}
~~~~~~Q_2 = (\bar su)_{V-A}\;(\bar ud)_{V-A} 
\end{equation}

{\bf QCD--Penguins:}
\begin{equation}\label{O2}
Q_3 = (\bar s d)_{V-A}\!\!\sum_{q=u,d,s,c,b}(\bar qq)_{V-A}~~~~~   
Q_4 = (\bar s_{\alpha} d_{\beta})_{V-A}\!\!\sum_{q=u,d,s,c,b}(\bar q_{\beta} 
       q_{\alpha})_{V-A} 
\end{equation}
\begin{equation}\label{O3}
Q_5 = (\bar s d)_{V-A}\!\!\sum_{q=u,d,s,c,b}(\bar qq)_{V+A}~~~~~
Q_6 = (\bar s_{\alpha} d_{\beta})_{V-A}\!\!\sum_{q=u,d,s,c,b}
      (\bar q_{\beta} q_{\alpha})_{V+A} 
\end{equation}

{\bf Electroweak Penguins:}
\begin{equation}\label{O4} 
Q_7 = \frac{3}{2}\,(\bar s d)_{V-A}\!\!\sum_{q=u,d,s,c,b} e_q\,(\bar qq)_{V+A} 
~~~~~Q_8 = \frac{3}{2}\,(\bar s_{\alpha} d_{\beta})_{V-A}\!\!\sum_{q=u,d,s,c,b}
      e_q\,(\bar q_{\beta} q_{\alpha})_{V+A}
\end{equation}
\begin{equation}\label{O5} 
 Q_9 = \frac{3}{2}\,(\bar s d)_{V-A}\!\!\sum_{q=u,d,s,c,b}e_q\,(\bar q q)_{V-A}
~~~~~Q_{10} =\frac{3}{2}\,
(\bar s_{\alpha} d_{\beta})_{V-A}\!\!\sum_{q=u,d,s,c,b}e_q\,
       (\bar q_{\beta}q_{\alpha})_{V-A} 
\end{equation}
Here, $\alpha,\beta$ denote colour indices and $e_q$ denotes the electric quark
charges reflecting the electroweak origin of $Q_7,\ldots,Q_{10}$. Finally,
$(\bar sd)_{V-A}\equiv \bar s_\alpha\gamma_\mu(1-\gamma_5) d_\alpha$. 

The Wilson coefficients $z_i$ and $y_i$ have been calculated at the NLO level
more than twenty years ago \cite{Buras:1993dy,Ciuchini:1993vr}, and some pieces
of NNLO corrections are also available \cite{Buras:1999st,Gorbahn:2004my,Brod:2010mj}. 
In Table \ref{tab:wcs}, we collect values for $z_{1,2}$ and $y_i$ at $\mu=m_c$,
used in our approach, for three values of $\alpha_s(M_Z)$ and $m_t=163\gev$,
in the  NDR-${\rm \overline{MS}}$  scheme.

\begin{table}[!htb]
\begin{center}
\begin{tabular}{|c|c|c|c|}
\hline
& $\alpha_s(M_Z)=0.1179$ & $\alpha_s(M_Z)=0.1185$ & $\alpha_s(M_Z)=0.1191$ \\
\hline
$z_1$ &   --0.4036 &   --0.4092 &   --0.4150 \\
$z_2$ & \phm1.2084 & \phm1.2120 & \phm1.2157 \\
\hline
$y_3$ & \phm0.0275 & \phm0.0280 & \phm0.0285 \\
$y_4$ &   --0.0555 &   --0.0563 &   --0.0571 \\
$y_5$ & \phm0.0054 & \phm0.0052 & \phm0.0050 \\
$y_6$ &   --0.0849 &   --0.0867 &   --0.0887 \\
\hline
$y_7/\al$ &      --0.0404 &   --0.0403 &   --0.0402 \\
$y_8/\al$ &    \phm0.1207 & \phm0.1234 & \phm0.1261 \\
$y_9/\al$ &      --1.3936 &   --1.3981 &   --1.4027 \\
$y_{10}/\al$ & \phm0.4997 & \phm0.5071 & \phm0.5146 \\
\hline
\end{tabular}
\end{center}
\caption{$\Delta S=1 $ Wilson coefficients at $\mu=m_c=1.3\gev$ for three
values of $\alpha_s(M_Z)$ and $m_t=163\gev$ in the  NDR-${\rm \overline{MS}}$  scheme.
\label{tab:wcs}}
\end{table}

\begin{boldmath}
\subsection{Basic formula for $\epe$}
\end{boldmath}
Our starting expression is formula (8.16) of \cite{Cirigliano:2003gt} which
we recall here in our notation\footnote{In order to simplify the notation
we denote $\RE(\epe)$ simply by $\epe$, which is real to an excellent
approximation. The latter is a model-independent consequence of the
experimentally known values of the (strong) phases of $\varepsilon'$ and
$\varepsilon$.}
\be\label{eprime0}
\frac{\varepsilon'}{\varepsilon} = -\,\frac{\omega_+}{\sqrt{2}\,|\varepsilon_K|}
\left[\, \frac{{\IM} A_0}{{\RE}A_0}\,(1-\Omega_{\rm eff}) -
\frac{{\IM} A_2}{{\RE}A_2} \,\right], 
\ee
where \cite{Cirigliano:2003gt}
\be\label{OM+}
\omega_+ =a\,\frac{\RE A_2}{\RE A_0} = (4.53\pm0.02)\times 10^{-2}, \quad
a = 1.017, \quad
\Omega_{\rm eff} = (6.0\pm 7.7)\times 10^{-2}\,.
\ee
Here $a$ and $\Omega_{\rm eff}$ summarise  isospin breaking corrections and
include  strong isospin violation $(m_u\neq m_d)$, the correction to the isospin
limit coming from $\Delta I=5/2$ transitions and electromagnetic corrections
\cite{Cirigliano:2003nn,Cirigliano:2003gt}. The amplitudes ${\RE}A_{0,2}$ are
then extracted from the branching ratios on $K\to\pi\pi$ decays in the isospin
limit. Their values are given in (\ref{eq:6.3}) below. In the limit $a=1$ and
$\Omega_{\rm eff}=0$ formula (\ref{eprime0}) reduces to the one used in
\cite{Bai:2015nea}, where all isospin breaking corrections except electroweak
penguin contributions have been set to zero.

The quantity $\Omega_{\rm eff}$ includes, in addition to other isospin breaking 
corrections, electroweak penguin contributions that are then not included 
in ${\IM}A_0$. Here we prefer to include these contributions to ${\IM}A_0$ and
therefore, instructed by the authors of \cite{Cirigliano:2003gt}, we remove
them from $\Omega_{\rm eff}$. However, we keep in $\Omega_{\rm eff}$ their
term $\Delta_0$ in the limit of $\alpha=0$. Using Table~4 of
\cite{Cirigliano:2003gt}, we then obtain the modified $\Omega_{\rm eff}$:
\be\label{OMeff}
\hat\Omega_{\rm eff} = (14.8 \pm 8.0)\times 10^{-2}\,.
\ee
As the second term in (\ref{eprime0}) is an isospin breaking effect by itself, 
strictly speaking in this term the parameter $a$ should be set to unity if 
we want to remove higher order isospin braking corrections. In addition, in
order to remove the effects of $\hat\Omega_{\rm eff}\not=0$ in electroweak
penguin contributions to ${\IM}A_0$, we write
\be\label{IMA0}
 {\IM}A_0=({\IM}A_0)^\text{QCDP}+ b\, ({\IM}A_0)^\text{EWP}, \qquad 
b=\frac{1}{a\,(1-\hat\Omega_{\rm eff})}
\ee
with the first term including the contributions from $Q_{3-6}$ and the 
second from  $Q_{7-10}$. Except for the tiny corrections due to $a\not=1$,
this procedure is equivalent to multiplying the coefficients $y_{3-6}$ by
$(1-\hat\Omega_{\rm eff})$ leaving $y_{7-10}$ unchanged.

Our final basic formula which we will use in what follows then reads
\be\label{eprime}
\frac{\varepsilon'}{\varepsilon} = -\,\frac{\omega_+}{\sqrt{2}\,|\varepsilon_K|}
\left[\, \frac{{\IM} A_0}{{\RE}A_0}\,(1-\hat\Omega_{\rm eff}) -
\frac{1}{a}\,\frac{{\IM} A_2}{{\RE}A_2} \,\right],
\ee
with $(\omega_+,a)$, $\hat\Omega_{\rm eff}$ and ${\IM}A_0$ given in (\ref{OM+}),
(\ref{OMeff}) and (\ref{IMA0}), respectively.  ${\IM}A_2$ contains only 
contributions of the electroweak penguin operators $Q_{7-10}$.

The crucial theory task for a precision SM prediction is to determine
the real and imaginary parts of the (strong-)isospin amplitudes
\be
A_I \equiv \langle (\pi \pi)_I |  {\cal H}_{\rm eff} | K \rangle
\ee
entering (\ref{eprime}) in terms of the Wilson coefficients and hadronic
matrix elements of the operators in the weak Hamiltonian (\ref{eq:Heff}).

\subsection{Hadronic matrix elements}

The hadronic matrix elements of the operators $Q_i$ entering the isospin
amplitudes,
\begin{equation}
\langle Q_i \rangle_I \equiv
\langle \left(\pi\pi\right)_I \left| Q_i \right| K \rangle \, ,
\qquad
I = 0,2 \, ,
\label{eq:5.1}
\end{equation}
generally depend on the scale $\mu$ and on the renormalisation scheme used for
the operators. These two dependencies are cancelled by those present in the
coefficients $C_i(\mu)$ so that the effective Hamiltonian and the resulting
amplitudes do not depend on $\mu$ and on the scheme used to renormalise the
operators. We will work exclusively in the NDR-${\rm \overline{MS}}$ scheme
and for scales $\mu\le m_c$, although in \cite{Buras:1993dy} also extensive
discussion of scales above $m_c$ can be found.

For $\mu\le m_c$, when the charm quark has been integrated out, only seven
of the operators listed above are independent of each other. Eliminating then
$Q_4$, $Q_9$ and $Q_{10}$ in terms of the remaining seven operators
results in the following important relations in the isospin limit
\cite{Buras:1993dy}:
\begin{eqnarray}
\langle Q_4 \rangle_0 &=&  \langle Q_3 \rangle_0 + \langle Q_2 \rangle_0
                          -\langle Q_1 \rangle_0 \,, \label{Q40}\\
\langle Q_9 \rangle_0 &=&
\frac{3}{2}\,\langle Q_1 \rangle_0 - \frac{1}{2}\,\langle Q_3 \rangle_0 \, ,
\label{eq:5.12} \\
\langle Q_{10} \rangle_0 &=&
    \langle Q_2 \rangle_0 + \frac{1}{2}\,\langle Q_1 \rangle_0
  - \frac{1}{2}\,\langle Q_3 \rangle_0 \, ,
\label{eq:5.13}\\
\langle Q_9 \rangle_2 &=&
   \langle Q_{10} \rangle_2 = \frac{3}{2}\,\langle Q_1 \rangle_2 \, ,
\label{eq:5.18}
\end{eqnarray}
where we have employed
\begin{eqnarray}
\langle Q_1 \rangle_2 &=&
\langle Q_2 \rangle_2  \, .
\label{eq:5.14} 
\end{eqnarray}

As stressed in \cite{Buras:1993dy}, in the NDR-${\rm \overline{MS}}$ scheme
the relation (\ref{Q40}) receives an $\ord(\alpha_s)$ correction due to the
presence of evanescent operators which have to be taken into account when using
Fierz identities in its derivation. The other relations above do not receive
such corrections. The complete expression for $\langle Q_4\rangle_0$ in the
NDR-${\rm \overline{MS}}$ scheme reads \cite{Buras:1993dy}
\begin{equation}
\langle Q_4 \rangle_0 = \langle Q_3 \rangle_0 + \langle Q_2 \rangle_0
                       -\langle Q_1 \rangle_0  -\frac{\alpha_s}{4\pi}
\Big( \langle Q_6\rangle_0 + \langle Q_4\rangle_0 - \frac{1}{3}\,
\langle Q_3\rangle_0 - \frac{1}{3}\,\langle Q_5\rangle_0 \Big) \,,
\label{eq:4.29b}
\end{equation}
which of course then has to be solved for $\langle Q_4\rangle_0$. However,
due to the partial cancellation between the matrix elements
$\langle Q_4\rangle_0$ and  $\langle Q_6\rangle_0$, and the smallness of the
matrix elements of $Q_3$ and $Q_5$, this correction affects the determination
of $\langle Q_4 \rangle_0$ by at most few percent and can be neglected. 
This procedure is supported both by the results on hadronic matrix elements 
RBC-UKQCD collaboration  \cite{Bai:2015nea} and the large-$N$ approach
\cite{Buras:2015xba}.

Setting the contribution of $Q_3$ to zero\footnote{In our numerical analysis
below, all operators will be taken into account.} and using the operators
\be
Q_\pm = \frac{1}{2} \big( Q_2\pm Q_1 \big) \,,
\ee
the formulae (\ref{Q40})-(\ref{eq:5.18}) read
\begin{eqnarray}
\langle Q_4 \rangle_0 &=&  2\,\langle Q_- \rangle_0 \,, \label{Q4a}\\
\langle Q_9 \rangle_0 &=&
\frac{3}{2}\,\big(\langle Q_+ \rangle_0 - \langle Q_- \rangle_0\big) \, ,
\label{eq:5.12a} \\
\langle Q_{10} \rangle_0 &=&
   \frac{3}{2}\,\langle Q_+ \rangle_0 + \frac{1}{2}\,\langle Q_-\rangle_0
   \, ,
\label{eq:5.13a}\\
\langle Q_9 \rangle_2 &=&
   \langle Q_{10} \rangle_2 = \frac{3}{2}\,\langle Q_+ \rangle_2 \, ,
\label{eq:5.18a}
\end{eqnarray}
which reduces the number of independent $(V-A) \otimes (V-A)$ matrix elements
entering ${\rm Re} A_{0,2}$ and ${\rm Im} A_{0,2}$ to three. On the other hand,
to an excellent approximation the amplitudes $\RE A_0$ and $\RE A_2$ at
$\mu=m_c$ are fully described by the operators $Q_-$ and $Q_+$, so that we can
write
\begin{eqnarray}\label{A0}
\RE A_0 &=& \frac{G_F}{\sqrt{2}}\, V_{ud} V_{us}^*\, \big( z_+\langle Q_+
\rangle_0+z_-\langle Q_- \rangle_0 \big) \,, \\
\label{A2}
\RE A_2 &=& \frac{G_F}{\sqrt{2}}\, V_{ud} V_{us}^* \,z_+\langle Q_+
\rangle_2 \,.
\end{eqnarray}

Introducing the ratio
\be\label{qratio}
q \equiv \frac{z_+(\mu)\langle Q_+(\mu) \rangle_0}{z_-(\mu)\langle Q_-(\mu)
\rangle_0} \,, \qquad z_\pm=z_2\pm z_1 \,,
\ee
allows us to express the ratios involving only $(V-A)\otimes (V-A)$ operators
that will enter our basic formula for $\epe$ as follows:
\begin{eqnarray}
\left(\frac{\IM A_0}{\RE A_0}\right)_{V-A} &=&
\IM\tau\,\frac{[4 y_4-b(3 y_9-y_{10})]}{2(1+q)z_-} +
\IM\tau\, b\, \frac{3q(y_9+y_{10})}{2(1+q)z_+} \,, \label{ratioa}\\
\left(\frac{\IM A_2}{\RE A_2}\right)_{V-A} &=&
\IM\tau\,\frac{3(y_9+y_{10})}{2z_+}\,.\label{ratiob}
\end{eqnarray}
Besides the CKM ratio $\tau$, the first ratio depends only on Wilson
coefficients and the single hadronic ratio $q$ to which we will return below.
On the other hand the second ratio is free from hadronic uncertainties, being
fully determined by the Wilson coefficients $z_+$, $y_9$, $y_{10}$ and by
$\tau$.

The remaining contributions to $\IM A_0$ and $\IM A_2$ are due to
$(V-A) \otimes (V+A)$ operators and are dominated by the operators
$Q_6$ and $Q_8$, respectively. We find this time
\be\label{ratio6}
\left(\frac{\IM A_0}{\RE A_0}\right)_6 = -\,\frac{G_F}{\sqrt{2}}\,
\IM\lambda_t\,y_6\,\frac{\langle Q_6 \rangle_0}{\RE A_0} \,,
\ee
\be\label{ratio8}
\left(\frac{\IM A_2}{\RE A_2}\right)_8 = -\,\frac{G_F}{\sqrt{2}}\,
\IM\lambda_t\,y^\text{eff}_8\,\frac{\langle Q_8 \rangle_2}{\RE A_2} \,.
\ee
Contributions from $Q_3$ and $Q_5$ are very suppressed but can and have been
included in our numerical error estimate. (See Appendix \ref{app:subleading}.)
We have also taken into account the small effect of $\langle Q_7 \rangle_2$,
for which a relatively precise lattice prediction exists \cite{Blum:2015ywa},
through the substitution
\be\label{y8eff}
y_8 \;\to\; y^\text{eff}_8 \equiv y_8 + p_{72}\, y_7 
\ee
which is included in writing (\ref{ratio8}). Here
$p_{72} \equiv \langle Q_7\rangle_2/\langle Q_8\rangle_2 =0.222$ for central
values of \cite{Blum:2015ywa}. (In our numerics, we have added the corresponding
errors linearly and attribute a 15\% uncertainty to this contribution.)

The matrix elements of the $Q_6$ and $Q_8$ operators are conveniently
parametrised by
\begin{eqnarray}\label{eq:Q60}
\langle Q_6(\mu) \rangle_0 &=& -\,4 h
\left[ \frac{m_{\rm K}^2}{m_s(\mu) + m_d(\mu)}\right]^2 (F_K-F_\pi)
\,B_6^{(1/2)} \,, \\
\label{eq:Q82}
\langle Q_8(\mu) \rangle_2 &=& \sqrt{2} h
\left[ \frac{m_{\rm K}^2}{m_s(\mu) + m_d(\mu)}\right]^2 F_\pi \,B_8^{(3/2)}\,,
\end{eqnarray}
with \cite{Buras:1985yx,Buras:1987wc}
\be\label{LN}
\bsi=\bei=1
\ee
in the large-$N$ limit. As had been demonstrated in \cite{Buras:1993dy},
$B_6^{(1/2)}$ and $B_8^{(3/2)}$ exhibit a very weak scale dependence.
The dimensionful parameters entering (\ref{eq:Q60}), (\ref{eq:Q82}) are given
by \cite{Agashe:2014kda,Aoki:2013ldr}
\be\label{FpFK}
m_K=497.614\mev, \qquad  F_\pi=130.41(20)\mev,\qquad \frac{F_K}{F_\pi}=1.194(5)\, ,
\ee
\be
m_s(m_c)=109.1(2.8)\mev, \qquad  m_d(m_c)=5.44(19)\mev\,.
\ee
In \cite{Aoki:2013ldr}, the light quark masses are presented at a scale of
$2\gev$, and we have evolved them to $\mu=m_c=1.3\gev$ with the help of the
renormalisation group equation. For the comparison with lattice results below,
we also need their values at $\mu=1.53\gev$, which are found to be
\be
m_s(1.53\gev)=102.3(2.7)\mev, \qquad  m_d(1.53\gev)=5.10(17)\mev\,.
\ee
Below, we will neglect the tiny errors on $m_K$, $F_K$, and $F_\pi$.

It should be emphasised that the overall factor $h$ in (\ref{eq:Q60}),
(\ref{eq:Q82}) depends on the normalisation of the amplitudes $A_{0,2}$.
In \cite{Buras:1993dy} and recent papers of the RBC-UKQCD collaboration
\cite{Blum:2012uk,Blum:2015ywa} $h=\sqrt{3/2}$ is used whereas in most recent
phenomenological papers
\cite{Cirigliano:2011ny,Buras:2014maa,Buras:2014sba,Buras:2015qea}, $h=1$. 
Correspondingly, the experimental values quoted for $A_{0,2}$ differ by this
factor. To facilitate comparison with \cite{Buras:1993dy} and the RBC-UKQCD
collaboration results \cite{Blum:2012uk,Blum:2015ywa,Bai:2015nea}, we will set
$h=\sqrt{3/2}$ in the present paper and consequently the experimental numbers
to be used are
\begin{equation}
\RE A_0 = 33.22(1)\times 10^{-8}\gev \, ,
\qquad \qquad
\RE A_2 = 1.479(3)\times 10^{-8}\gev \, ,
\label{eq:6.3}
\end{equation}
which display the $\Delta I=1/2$ rule
\begin{equation}
\frac{\RE A_0}{\RE A_2}\equiv \frac{1}{\omega} = 22.46 \,.
\label{eq:6.4}
\end{equation}
We also note that while equation (\ref{eq:Q60}) is identical to (5.10) in
\cite{Buras:1993dy}, the definition of $B_8^{(3/2)}$ in the present paper
differs from \cite{Buras:1993dy} [cf (5.18) there]. This is to ensure that
$\bsi=1$ and $\bei=1$ both correctly reproduce the large-$N$ limit of QCD.
In contrast, (5.18) in \cite{Buras:1993dy} was based on the so-called vacuum
insertion approximation, in which additional terms appear in the normalisation
of $B_8^{(3/2)}$. Such terms misrepresent the large-$N$ limit of QCD. With our
conventions, $1/N$ corrections in (\ref{eq:Q60}) and (\ref{eq:Q82}) are
represented by the departure of $\bsi$ and $\bei$ from unity. They have been
investigated in \cite{Hambye:1998sma} and very recently in \cite{Buras:2015xba} with
the result summarised in (\ref{NBOUND}). We refer to this paper for further 
details.

We now turn to the parameter $q$ which enters (\ref{ratioa}). We first note
that, like $\bsi$ and $\bei$, it is nearly renormalisation-scale independent.
Its value can be estimated in the large-$N$ approach \cite{Buras:2014maa}; as
this approach correctly accounts for the bulk of the experimental value of
${\rm Re}A_0$, the ensuing estimate can be considered a plausible one. In the
large-$N$ limit, corresponding to $\mu=0$, one finds first
$\langle Q_+(0)\rangle_0/\langle Q_-(0)\rangle_0=1/3$. Using the meson
evolution in \cite{Buras:2014maa} up to $\mu=1.0\gev$ and then quark evolution
up to $\mu=m_c$, multiplying the result by $z_+(m_c)/z_-(m_c)$, we obtain
$q\approx 0.1$. On the other hand the results of the RBC-UKQCD collaboration
\cite{Bai:2015nea} are consistent with a value of zero
($q=0.029\pm 0.087$). As the large-$N$ approach gives ${\rm Re}A_0$ below
the data while \cite{Bai:2015nea} above it, we expect the true value of $q$ at
$\mu=m_c$ to lie between these two estimates and will take $q$ in the range
\be  \label{eq:qrange}
0\le q \le 0.1 \,.
\ee
We consider this a credible range, but already mention that our phenomenological
results below would change very little even if we enlarged this range by a
factor of a few: $q$ is simply too small to introduce a large error on $\epe$.

Our input parameters including sub-leading hadronic parameters defined in
Appendix \ref{app:subleading}  are collected in Table \ref{tab:inputs}.
Regarding ${\rm Im} \lambda_t$, we choose a central value between the UTfit
\cite{UTfit} and CKMfitter \cite{Charles:2015gya} determinations and an error
slightly larger than that obtained from either fit. This is to account for the
very small errors on $V_{ud}$ and $V_{us}$, which we fix to PDG central values
\cite{Agashe:2014kda}. The Wilson coefficients in Table \ref{tab:wcs} come with
an additional uncertainty from unknown higher-order corrections. In particular
the threshold corrections at $m_c$ can be substantial even at NNLO. This
can for example be seen in the perturbative convergence of $\varepsilon_K$
\cite{Brod:2010mj,Brod:2011ty}. We use a scale variation to establish the
typical size of higher order corrections and estimate a 10\% uncertainty for
each Wilson coefficient $y_3$ -- $y_{10}$ of Table \ref{tab:wcs}.

\begin{table}[!htb]
\begin{center}
\begin{tabular}{|c|c|c|}
\hline
 & value range & comment \\
\hline
$B_6^{(1/2)}$ & $0.57 \pm 0.19$  & Eq.\ (\ref{Lbsi}) and surrounding discussion \\
$B_8^{(3/2)}$ & $0.76 \pm 0.05$ & Eq.\ (\ref{Lbei}) and surrounding discussion\\
$q$ & $0.05 \pm 0.05$ & see (\ref{qratio}), (\ref{eq:qrange}) \\
$B_8^{(1/2)}$ & $1.0 \pm 0.2$ & defined in Eq.\ (\ref{eq:Q80}) \\
$p_{72}$ & $0.222 \pm 0.033$  & Eq.\ (\ref{y8eff}) and surrounding discussion \\
$p_3$ & $ 0 \pm 0.5$ & see Appendix \ref{app:subleading} \\
$p_5$ & $0 \pm 0.5$ & see Appendix \ref{app:subleading}  \\
$p_{70}$ & $0 \pm 1/3$ & see Appendix \ref{app:subleading} \\
\hline
${\rm Im} \lambda_t$ & $(1.4 \pm 0.1) \times 10^{-4}$ & see text \\
$m_t(m_t)$ & $ (163 \pm 3)$ GeV &  calculated from pole mass value \cite{Agashe:2014kda} \\
$m_s(m_c)$ & $(109.1 \pm 2.8)$ GeV & value from \cite{Agashe:2014kda}, evolved \\
$m_d(m_c)$ & $(5.4 \pm 1.9)$ GeV & value from \cite{Agashe:2014kda},
evolved \\
$\alpha_s(M_Z)$ & $0.1185 \pm 0.0006$ & from \cite{Agashe:2014kda} \\
$s^2_W$ & $0.23126$ & $\overline{\mathrm{MS}}$ scheme value from \cite{Agashe:2014kda} \\
\hline
$\hat\Omega_{\rm eff}$ & $(14.8 \pm 8.0) \times 10^{-2} $ &from \cite{Cirigliano:2003gt}  \\
\hline
$y_{3}$ -- $y_{10}$ & $y_{i} \times (1 \pm 0.1)$ &  see Text \\
\hline
\end{tabular}
\end{center}
\caption{Input parameter ranges, grouped into: hadronic matrix elements,
 parametric, isospin breaking and NNLO. The (numerically unimportant) ratios
 $p_{72}$, $p_3$, $p_5$, $p_{70}$ are defined in Appendix \ref{app:subleading}).
 The remaining parameters ($F_\pi$, $F_K$, $m_K$, $V_{ud}$, $V_{us}$,
 $\alpha_{\rm em}$, $G_F$, $\varepsilon_K$) are fixed at their central values.
\label{tab:inputs}}
\end{table}

\begin{boldmath}
\subsection{Convenient formula for $\epe$}
\end{boldmath}

Before turning to quantitative phenomenology, in order to make easier
connection with the phenomenological literature and aid discussion of our
results, we summarise the discussion so far in a concise formula (derived
first in \cite{Buras:1993dy}) for $\epe$ that exhibits the sensitivity to the
two most important hadronic matrix elements $\bsi$ and $\bei$ transparently.

Using the effective Hamiltonian (\ref{eq:Heff}) and the experimental data
for $\omega$, $\RE A_0$ and $\varepsilon_K$, we find 
\begin{equation}
\frac{\varepsilon'}{\varepsilon} = \, \IM\lambda_{\rm t} \cdot
\left[\, a \big(1-\hat\Omega_{\rm eff}\big)\,P^{(1/2)}-P^{(3/2)}\,\right] \, ,
\label{eq:8.3}
\end{equation}
where  
\begin{eqnarray}
P^{(1/2)} & = & \sum P^{(1/2)}_i \; = \;
              r \sum y_i \langle Q_i
\rangle_0 \, , \label{eq:8.4} \\
P^{(3/2)} & = & \sum P^{(3/2)}_i \; = \;
\frac{r}{\omega} \sum y_i \langle Q_i \rangle_2 \, , \label{eq:8.5}
\end{eqnarray}
with
\begin{equation}
r \, = \, \frac{G_F\,\omega}{2\,|\varepsilon_K|\,\RE A_0}\,.
\label{eq:8.6}
\end{equation}
In (\ref{eq:8.4}) and (\ref{eq:8.5}) the sums run over all contributing
operators. Therefore in $P^{(1/2)}$ in the case of EWP contributions we have
to take into account the correction $b\not=1$ defined in (\ref{IMA0}).

Writing then
\begin{eqnarray}
P^{(1/2)} & = & a_0^{(1/2)}  +
                  a_6^{(1/2)}\,B_6^{(1/2)} \, ,
\label{eq:8.10} \\
\mvs
P^{(3/2)} & = & a_0^{(3/2)} + a_8^{(3/2)}\,B_8^{(3/2)} \, ,
\label{eq:8.11}
\end{eqnarray}
with the parameters $\bsi$ and $\bei$ taken at $\mu=m_c$ and using the 
expressions (\ref{ratioa})-(\ref{eq:Q82}) we find:
\begin{eqnarray}
a_0^{(1/2)}  &=& r_1 \biggl[\, \frac{[4 y_4-b(3 y_9-y_{10})]}{2(1+q)z_-} +
b\,\frac{3q(y_9+y_{10})}{2(1+q)z_+} \,\biggr] + r_2\,b\, y_8 \,
\frac{\langle Q_8 \rangle_0}{\RE A_0} \,, \label{a012} \\
a_6^{(1/2)} &=& r_2\, y_6 \, \frac{\langle Q_6 \rangle_0}{\bsi
\RE A_0} \,, \label{a06}\\
a_0^{(3/2)} &=& r_1\, \frac{3(y_9+y_{10})}{2z_+} \,, \label{a032} \\
a_8^{(3/2)} &=& r_2\, y^\text{eff}_8 \,
\frac{\langle Q_8 \rangle_2} {\bei \RE A_2} \,, \label{a28}
\end{eqnarray}
where 
\be
r_1=\frac{\omega}{\sqrt{2}|\varepsilon_K|}\frac{1}{V_{ud} V_{us}^*}\,, \qquad
r_2=\frac{\omega}{2|\varepsilon_K|} G_F \,,
\ee
and $\langle Q_6\rangle_0$, $\langle Q_8\rangle_0$, and $\langle Q_8\rangle_2$
are given in (\ref{eq:Q60}), (\ref{eq:Q80}) and (\ref{eq:Q82}), respectively.
The second term in (\ref{a012}) proportional to $q$ amounts at most to a 2\%
correction and could be safely neglected. $y^\text{eff}_8$ is defined in
(\ref{y8eff}). $a_0^{(1/2)}$ and $a_0^{(3/2)}$ receive further small corrections
which can be extracted from the expressions in Appendix \ref{app:subleading}.
Apart from that, the coefficients $a_i^{(1/2)}$ and  $a_i^{(3/2)}$ depend only
on $q$, $\alpha_s$, $m_t$, and the renormalisation scheme considered. The
dependencies on $\alpha_s$ and $m_t$ are given in the  NDR-${\rm \overline{MS}}$
scheme in Table~\ref{tab:ais}.

\begin{table}[!htb]
\begin{center}
\begin{tabular}{|c|c||c|c|c|c|}
\hline
$\alpha_s(M_Z)$ & $m_t$ [GeV] & $a_0^{(1/2)}$ & $a_6^{(1/2)}$ &
$a_0^{(3/2)}$ & $a_8^{(3/2)}$ \\
\hline
       & 160 & $-\,2.93(12)$ & 17.23 & $-\,0.82$ & 6.96 \\
0.1179 & 163 & $-\,2.90(12)$ & 17.25 & $-\,0.84$ & 7.27 \\
       & 166 & $-\,2.87(12)$ & 17.26 & $-\,0.85$ & 7.58 \\
\hline
       & 160 & $-\,2.95(12)$ & 17.61 & $-\,0.82$ & 7.13 \\
0.1185 & 163 & $-\,2.92(12)$ & 17.63 & $-\,0.84$ & 7.44 \\
       & 166 & $-\,2.89(12)$ & 17.64 & $-\,0.85$ & 7.76 \\
\hline
       & 160 & $-\,2.98(12)$ & 18.00 & $-\,0.82$ & 7.31 \\
0.1191 & 163 & $-\,2.95(12)$ & 18.02 & $-\,0.84$ & 7.62 \\
       & 166 & $-\,2.92(12)$ & 18.03 & $-\,0.85$ & 7.95 \\
\hline
\end{tabular}
\end{center}
\caption{The coefficients $a_i^{(1/2)}$ and $a_i^{(3/2)}$ in the
NDR-${\rm \overline{MS}}$  scheme for different values of $\alpha_s(M_Z)$
and $m_t$. The uncertainty shown for $a_0^{(1/2)}$ only includes the variation
of $q$.\label{tab:ais}}
\end{table}

In summary the ratio $\epe$ is governed by the following four contributions:
\begin{itemize}
\item[i)]
The contribution of $(V-A)\otimes (V-A)$ operators to $P^{(1/2)}$ is represented
by the first term in (\ref{eq:8.10}). As seen in (\ref{a012}) this term is
governed by the operator $Q_4$ and includes also small contributions from
$(V-A)\otimes (V-A)$ electroweak penguin operators. We find that this term is
{\em negative} and only weakly dependent on $q$. Also the dependences on
$\alpha_s$ and renormalisation scheme (see \cite{Buras:1993dy}) are weak.
These weak dependences originate from the fact that in our approach the matrix
elements entering the first term in $P^{(1/2)}$ cancel out. The weak
dependence on $m_t$ results from the contributions of sub-leading electroweak
penguin operators and is exhibited in the formulae in Appendix~\ref{app:X}.
As pointed out in \cite{Buras:1993dy}, the suppression of $\epe$ through
$a_0^{(1/2)}$ increases with increasing $\RE A_0$, a feature which in the next
section will help us to partly understand the result in (\ref{RBC}).
\item[ii)]
The contribution of $(V-A)\otimes (V+A)$ QCD penguin operators to $P^{(1/2)}$
is given by the second term in (\ref{eq:8.10}). This contribution is large
and {\em positive} and is dominated by the operator $Q_6$. The coefficient
$a_6^{(1/2)}$ depends sensitively on $\alpha_s$, but as in the last two decades
the precision on $\alpha_s$ increased, this uncertainty is small in 2015 as
can be seen from Table~\ref{tab:ais}.
\item[iii)]
The contribution of the $(V-A)\otimes (V-A)$ electroweak penguin operators
$Q_9$ and $Q_{10}$ to $P^{(3/2)}$ is represented by the first term in
$P^{(3/2)}$. As in the case of the contribution i), the matrix elements
contributing to $a_0^{(3/2)}$ cancel out in the SM. Consequently, the scheme
and $\alpha_s$ dependences of $a_0^{(3/2)}$ are weak. As seen in (\ref{a032})
the sizable $m_t$-dependence of $a_0^{(3/2)}$ results from the corresponding
dependence of $y_9 + y_{10}$ but again the precision on $m_t$ increased by
much in the last two decades.  $a_0^{(3/2)}$ contributes {\em positively} to
$\epe$. 
\item[iv)]
The contribution of the $(V-A)\otimes (V+A)$ electroweak penguin operators
$Q_7$ and $Q_{8}$ to $P^{(3/2)}$ is represented by the second term in
(\ref{eq:8.11}). This contribution is dominated by $Q_8$ and depends sensitively
on $m_t$ and $\alpha_s$. It contributes {\em negatively} to $\epe$. 
\end{itemize}
The competition between these four contributions is the reason why it is
difficult to predict $\epe$ precisely. In this context, one should appreciate
the virtue of our approach: the contributions i) and iii) can be determined
rather precisely by CP-conserving data so that the dominant uncertainty in our
approach in predicting $\epe$ resides in the values of $\bsi$ and $\bei$.

\section{Prediction for \boldmath{$\epe$} in the SM}\label{sec:2a}
\begin{boldmath}
\subsection{Prediction for $\epe$ and discussion}
\end{boldmath}

We begin our analysis by employing the lattice values in (\ref{Lbei}) and
(\ref{Lbsi}). Varying all parameters within their input ranges and combining
the resulting variations in $\epe$ in quadrature, we obtain:
\be\label{LATeps}
    (\epe)_\text{SM}  = (1.9 \pm 4.5) \times 10^{-4} .
\ee
Comparing to the experimental result
$(\epe)_{\rm exp} = (16.6 \pm 2.3) \times 10^{-4}$ (average of NA48
\cite{Batley:2002gn} and KTeV \cite{AlaviHarati:2002ye,Worcester:2009qt}), we
observe a discrepancy of $2.9\,\sigma$ significance.

\begin{table}[!htb]
\begin{center}
\begin{tabular}{|c|c||c|c|}
\hline
quantity  & error on $\epe$ & quantity & error on $\epe$ \\
\hline
$B_6^{(1/2)}$ & $4.1$ &            $m_d(m_c)$ & $0.2$ \\
NNLO & $1.6$ &                     $q$ & $0.2$ \\
$\hat{\Omega}_{\rm eff}$ & $0.7$ & $B_8^{(1/2)}$ & $0.1$ \\
$p_3$ &    $0.6$ &                 ${\rm Im} \lambda_t$ & $0.1$ \\
$B_8^{(3/2)}$ & $0.5$ &            $p_{72}$ & $0.1$ \\
$p_5$ &      $0.4$ &               $p_{70}$ & $0.1$\\
$m_s(m_c)$ & $0.3$ &               $\alpha_s(M_Z)$ & $0.1$ \\
$m_t(m_t)$ & $0.3$ &               & \\
\hline
\end{tabular}
\end{center}
\caption{Error budget, ordered from most important to least important.
 Each line shows the  variation from the central value of our $\epe$
 prediction, in units of $10^{-4}$, as the corresponding parameter is varied
 within its input range, all others held at central values.
\label{tab:errbud}}
\end{table}

A detailed error budget is given in Table \ref{tab:errbud}. It is evident that
the error is dominated by the hadronic parameter $\bsi$. Uncertainties from
higher-order corrections are still significant yet small if compared to the
deviation from the experimental value. All other individual errors are below
$10^{-4}$, with the third most important uncertainty coming from the isospin
breaking parameter $\hat\Omega_{\rm eff}$, at a level of $0.7 \times 10^{-4}$
and about six times smaller than the error due to $\bsi$. If matrix elements
are taken from a lattice calculation, the $m_s$ dependence is only an artifact
of our parametrisation in terms of $\bsi$ and $\bei$. Therefore including the
$m_s$ variation in our error estimate for $\epe$ leads to a slight, but
negligible, overestimate of the total error. At the same time, the small $m_s$
dependence we do find in the final result shows that this is no longer a
relevant source of uncertainty in non-lattice approaches (like the large-$N$
approach in particular) in which $\bsi$ and $\bei$ are directly calculated.

At this stage it is important to emphasise that the results for $\bei$ and 
$\bsi$ in (\ref{Lbei}) and (\ref{Lbsi}) receive strong support from the
large-$N$ approach as recently demonstrated in \cite{Buras:2015xba}. In
particular the smallness of the matrix element $\langle Q_6\rangle_0$ with
respect to $\langle Q_8\rangle_2$ is the result of the chiral suppression of
$\langle Q_6\rangle_0$, signalled by $F_K-F_\pi$ in (\ref{eq:Q60}). As seen
in (\ref{eq:Q82}) no such suppression is present in $\langle Q_8\rangle_2$.
But in addition it is possible to demonstrate that both $\bsi$ and $\bei$ are
below unity as given in (\ref{NBOUND}). Moreover, while $\bei=0.8\pm 0.1$ is
found in this approach, the values of $\bsi$ are in the ballpark of the lattice
result and consequently give a strong support for $\bsi < \bei$  as indicated
by the lattice data. But as present calculations by lattice QCD and in
\cite{Buras:2015xba} are not precise enough, at this moment, we cannot exclude
that $\bsi$ could be as large as $\bei$ and this leads conservatively to the
bound in (\ref{NBOUND}).

For these reasons it is instructive to consider other values of the parameters
$\bsi$ and $\bei$ than those obtained by RBC-UKQCD collaboration which are,
however, consistent with the large-$N$ bound in (\ref{NBOUND}). Of particular
interest is the choice $\bsi=\bei =1$ which corresponds to the saturation of
this bound and the choice in which the bound on $\bsi$ is saturated when $\bei$
is fixed to the central lattice value in (\ref{Lbei}). Using the same input for
the remaining parameters, we find
\be\label{eps1}
   (\epe)_\text{SM} = (8.6 \pm 3.2) \times 10^{-4}, \qquad (\bsi=\bei=1),
\ee
\be\label{eps2}
   (\epe)_\text{SM} = (6.0 \pm 2.4) \times 10^{-4}, \qquad (\bsi=\bei= 0.76).
\ee
We observe that even for these values of $\bsi$ and $\bei$ the SM predictions
for $\epe$ are significantly below the data. This is an important result as it 
shows that even if the value of $\bsi$ from lattice calculations would move up 
in the future, the SM would face difficulty in reproducing the data provided
the large-$N$ bound in  (\ref{NBOUND}) is respected.

With these results at hand, we are in the position to summarise the present
picture of the estimate of $\epe$ in the SM:
\begin{itemize}
\item
First, parametric uncertainties decreased by much since the analyses of $\epe$
around the year 2000. This includes the uncertainty in ${\rm Im} \lambda_t$
which is presently about $\pm\,7\%$ and is irrelevant in the estimate in
(\ref{LATeps}) but plays some role when $\epe$ is larger. Also the improvement
on $m_s$ should be appreciated, entailing that the uncertainty on $m_s$ no
longer is an issue.
\item
Second, the previously sizeable uncertainty due to $\bei$ has become
sub-dominant, much smaller for example than the one due to isospin violation.
This is thanks to impressive progress on the lattice \cite{Blum:2015ywa}, which
confirms large-$N$ estimates employed in our previous papers, but with far
smaller uncertainty.
\item
Third, {the present analysis further increased the effectiveness of our 
framework,} leading to a situation in which a single parameter $\bsi$
is playing the decisive role in the answer to the question whether $\epe$ in
the SM can be reconciled  with the data or not. The new finding both by the
lattice QCD and large-$N$ approach that $\bsi$ is below unity narrowed
significantly the range for $\epe$ in the SM in our framework.
\end{itemize}

This picture clearly indicates the emergence of a new anomaly in $K$ physics.
As this anomaly is strictly correlated in our framework with the value of
$\bsi$, this parameter  must be a priority for future non-perturbative
calculations for flavour physics. Fortunately, it is accessible by
first-principle lattice-QCD calculations. Systematic improvement is hence
possible. (See also comparison with lattice below.) Progress on isospin
violation will also be important. 

But already now, the results presented here  motivate further scrutiny of the
SM prediction as well as searching for viable beyond-SM explanations. We will
briefly discuss both directions, in Sections \ref{sec:explSM} and
\ref{sec:explBSM}, respectively.

Last but not least, the great reduction in parametric and hadronic
uncertainties, made effective through our formalism, and good prospects on
$\bsi$, may make a more precise measurement of $\epe$ in the future worthwhile.

\begin{boldmath}
\subsection{Discussion of $\bsi$ dependence}
\end{boldmath}

The domination of our error estimate by the uncertainty on $\bsi$ leads
us to investigate the dependence of $\epe$ on $\bsi$ in more detail.

There is a hierarchy in the four contributions  discussed in the previous
section with ii) being most important followed by iv), i) and iii). For central
values of {input parameters} we find
\begin{equation}
\frac{\varepsilon'}{\varepsilon} = 10^{-4} \biggl[
\frac{\IM\lambda_{\rm t}}{1.4\cdot 10^{-4}}\biggr]\!\left[\,a\,
\big(1-\hat\Omega_{\rm eff}\big) \big(-4.1(8) + 24.7\,\bsi\big) + 1.2(1) -
10.4\,\bei \,\right],
\label{AN2015}
\end{equation}
with the four terms corresponding to the four contributions in question.
The first number in brackets comprise the uncertainties of the sub-leading
hadronic parameters $q$, $p_3$, $p_5$, $p_{70}$ and $B_8^{(1/2)}$, while the
second number in brackets is due to the uncertainty in $p_{72}$. This
assignment of uncertainties will simplify the comparison with (\ref{ANLAT}),
even though it does not strictly follow our formalism. Furthermore, a remark
on error correlations is in order. Due to implementing the constraints from
CP-conserving data, correlations between the different contributions to $\epe$
are introduced. However, as the initial correlations of the hadronic matrix
elements determined on the lattice are not available, we refrain from
incorporating them into our analysis.

It should be noted that the term representing $Q_6$ penguin operator involves 
the product $a(1-\hat\Omega_{\rm eff})\bsi$. Therefore, effectively isospin 
breaking corrections lower the value of $\bsi$ by $0.866$, implying in 
the case of $\bsi=0.57$ an effective value of $0.49$.

\begin{figure}[!htb]
\begin{center}
\includegraphics[height=8cm]{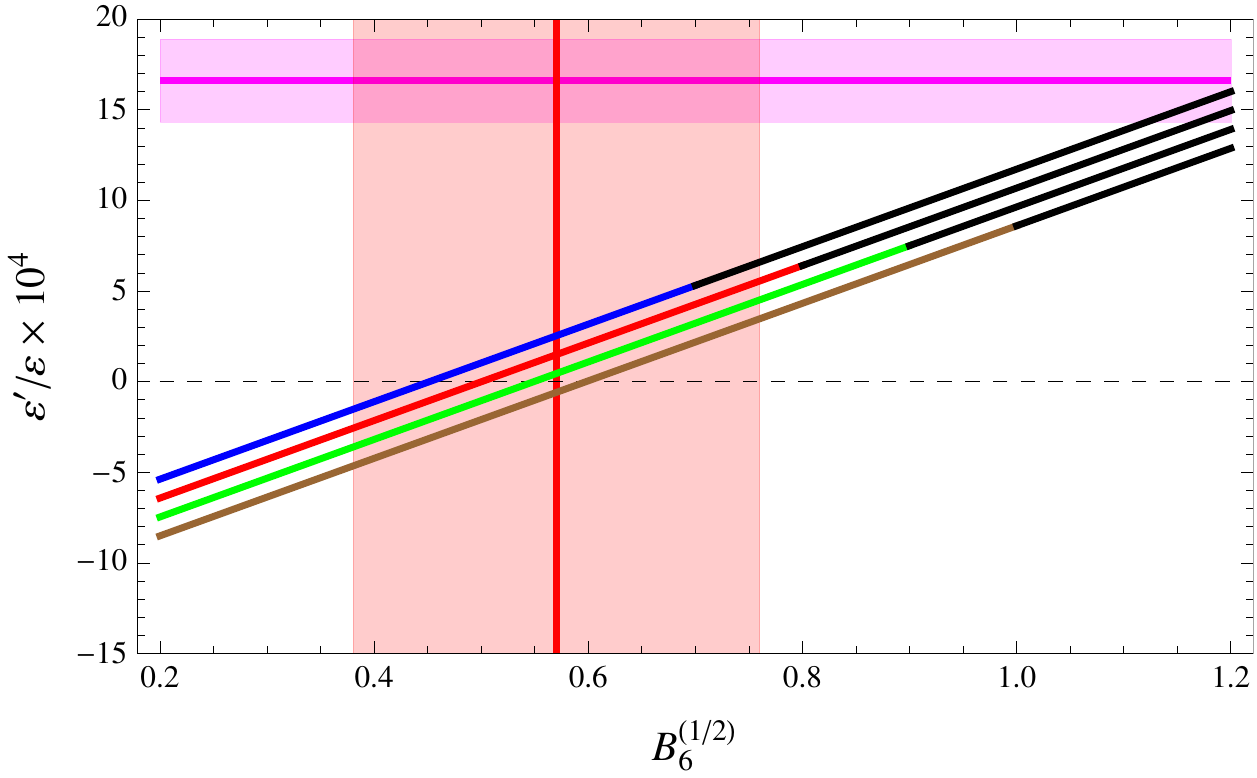}
\caption{$\epe$ as a function of $\bsi$. For further explanation see the text.
\label{fig:epeb6b8} }
\end{center}
\end{figure}

In Figure~\ref{fig:epeb6b8}, we show $\epe$ as a function of $\bsi$ for
different values of $\bei$: 
\begin{eqnarray}
\bei &=& 0.7~(\text{blue}), \quad \bei=0.8~(\text{red}), \\
\bei &=& 0.9~(\text{green}), \quad \bei=1~(\text{brown})\,.
\end{eqnarray}
The vertical band represents central value and error on $\bsi$ from
(\ref{Lbsi}), the horizontal band the experimental world average on $\epe$.
The black region on each line is excluded by the bound (\ref{NBOUND}).
We observe that the experimental value of $\epe$ can only be reproduced in the
SM far outside the RBC-UKQCD range and then only for values $\bsi > \bei$ and
$\bsi> 1$ in variance with the bound (\ref{NBOUND}).

We finally observe that even if the bound $\bsi\le \bei$ is violated, but the
bound $\bsi\le 1$ is respected, the SM cannot quite reach the experimental
data. Indeed, employing this unlikely hypothesis, we find this time
\be\label{eps3}
   (\epe)_\text{SM} = (11.1 \pm 3.2) \times 10^{-4}, \qquad
   (\bsi=1.0,\quad \bei= 0.76).
\ee


\section{Comparison with RBC-UKQCD lattice QCD}\label{sec:3}

\subsection{Preliminaries}

The results for $\epe$ presented in \cite{Blum:2015ywa,Bai:2015nea} can be
summarised by a formula analogous to (\ref{AN2015}),
\begin{equation}
\frac{\varepsilon'}{\varepsilon} = \,10^{-4}\,\biggl[\frac{\IM\lambda_{\rm t}}
{1.4\cdot 10^{-4}}\biggr] \left[\,-\,6.5(3.2) + 25.3\, \bsi + 1.2(8) -
10.2\,\bei \right] .
\label{ANLAT}
\end{equation}

In deriving this formula, we used the value of the matrix element
$\langle Q_6\rangle$  given in \cite{Bai:2015nea} for $\mu=1.53\gev$:
\be
\langle Q_6(\mu) \rangle_0 = -\,0.379(97)(83)\gev^3 \qquad (\mbox{RBC-UKQCD}) \,
\ee 
with the first error being statistical and the second systematic.
Using (\ref{eq:Q60}), we find (see also \cite{Buras:2015xba})
\be\label{Lbsi0}
B_6^{(1/2)}(\mu=1.53\gev)=0.57\pm 0.19\,, 
\ee
and consequently (\ref{Lbsi}). The value of $\bsi$ is significantly lower
than its upper limit from the large-$N$ approach in (\ref{NBOUND}) \cite{Buras:2015xba}
and the values for $\bsi$ used in many papers until now. This is the central
reason why the lattice result is substantially below the data.

Using (\ref{eq:Q82}) and comparing to the corresponding matrix element in 
\cite{Blum:2015ywa} one extracts \cite{Buras:2015qea}
\be\label{Lbei0}
B_8^{(3/2)}(3\gev)=0.75\pm 0.05, \qquad B_8^{(3/2)}(m_c)=0.76\pm 0.05,
\ee
which displays the very weak $\mu$ dependence mentioned above.

Setting $\bsi= 0.57\pm 0.19$ and $\bei=0.76\pm 0.05$, we indeed obtain the
result in (\ref{RBC}).

Comparing formulae (\ref{AN2015}) and (\ref{ANLAT}), we observe the following 
differences: 
\begin{itemize}
\item
In \cite{Bai:2015nea}, $a=1$ and $\hat\Omega_{\rm eff}=0$  have been employed.
\item
The main difference for fixed $\bsi$ and $\bei$ is found in the first term in
(\ref{ANLAT}). Not only is the error in this term much larger than in our 
formula but also is this term significantly larger than found by us.
\item
Also striking is the sizeable error in the third term which is very small in 
our case.
\end{itemize}

Let us then have a closer look at the contribution of the $Q_4$ operator in
order to clarify the reason for this difference.

\begin{boldmath}
\subsection{Contribution of $Q_4$ and $\RE A_0$}
\end{boldmath}

Using the formulae of the previous section, we readily find 
\be\label{Q4}
\langle Q_4 (m_c)\rangle_0 = \frac{2 \sqrt{2}}{(1+q)z_-}\,
\frac{\RE A_0}{G_F V_{ud} V_{us}^*} \,.
\ee
For $q=0.05$, using the experimental value of $\RE A_0$, we obtain
\be\label{Q4BJL}
\langle Q_4(m_c) \rangle_0 = 0.22(1)\gev^3. 
\ee
On the other hand in \cite{Bai:2015nea} $q\approx 0$ and 
\be
\RE A_0= 4.62(0.95)(0.27)\times 10^{-7}\gev,
\ee
the central value of which is roughly $40\%$ larger than the experimental
value in (\ref{eq:6.3}). From (\ref{Q4}) we now find
\be\label{Q4BJLL}
\langle Q_4(m_c) \rangle_0 = 0.31(7)\gev^3\,.
\ee

This value agrees with the one given in \cite{Bai:2015nea}:
\be\label{Q4LATTICE}
\langle Q_4(1.53\gev) \rangle_0 = 0.271(93)(60)\gev^3 \qquad (\mbox{RBC-UKQCD}).
\ee

But what is striking is the high precision obtained for this matrix element 
in our approach and still large uncertainty in the lattice result. It
should also be noted that the contribution of the $Q_4$ operator to $\epe$ is
in the present lattice result comparable to the one of $Q_8$ and can be even
larger than the latter one, which is not possible in our approach.

\subsection{Electroweak contribution}

On the other hand the electroweak penguin contribution to $\epe$ is similar
because the lattice value for $\RE A_2$ agrees well with experiment
\cite{Blum:2015ywa}. Using the lattice result $\bei=0.76\pm0.05$, we find
\be\label{EWPus}
(\epe)_\text{EWP}= -\,(6.7 \pm 0.5)\times 10^{-4} \,,
\ee
which can also be obtained from the last two terms in (\ref{AN2015}). This 
result compares well with  \cite{Blum:2015ywa}
\be
(\epe)_\text{EWP}= -\,(6.6\pm 1.0)\times 10^{-4}\,,\qquad (\mbox{RBC-UKQCD}),
\ee
although our error is substantially smaller.


\section{Can the large observed \boldmath{$\epe$} be made consistent with
the SM?}\label{sec:explSM}

Given the significant discrepancy between our SM prediction and the
experimental result, we first consider possible missing or
underestimated contributions in the SM.

\subsection{Missing chromomagnetic contributions}

In our discussion (and much of the literature) the chromomagnetic penguin
$Q_{8g}$ (and also its electromagnetic counterpart $Q_{7\gamma}$) have been
tacitly dropped. It is straightforward to extend the formalism to include
$Q_{8g}$, which being pure $\Delta I=1/2$ impacts only on ${\rm Im} A_0$.
While $y_{7\gamma}$ is small compared to the leading electroweak penguin
coefficients, precluding any effect, the coefficient $y_{8g}$ is sizeable.
The status of the hadronic matrix element $\langle Q_{8g} \rangle_0$ is rather
uncertain. A calculation at leading non-vanishing order in the chiral quark
model \cite{Bertolini:1993rc} gave
\be  \label{eq:Q8MELT}
  \langle Q_{8g} \rangle_0 = -\,h\, \frac{1}{16\,\pi^2} \frac{11}{2}
  \frac{m_s}{m_s+m_d} \frac{F_K^2}{F_\pi^3}\, m_K^2 m_\pi^2\, B_{8g} \,,
\ee
(recall $h=\sqrt{3/2}$ in our normalisation) with $B_{8g}=1$, obtaining an
upward shift of about  $0.3 \times 10^{-4}$ on $\epe$. Due to uncertainties
from unknown higher orders and $1/N$ corrections, an ad-hoc range $1 \leq
B_{8g} \leq 4$ was advocated in \cite{Buras:1999da} for setting bounds
on new physics. For $C_{8g}(m_c) \approx -\,0.185$ and central values of
our other input parameters, the resultant shift is in the range
\be
   \Delta \left. \frac{\varepsilon'}{\varepsilon} \right|_{Q_{8g}} = (0.2 \dots 0.7) \times 10^{-4} .
\ee
At the upper end of the range, while still being insufficient to explain
the tension between theory and experiment, the contribution becomes competitive
with some of the larger sub-leading uncertainties. Although a chromomagnetic
contribution has never been seriously considered as a sizable SM contribution,
the possibility cannot be fully excluded. A more definite conclusion would be
desirable and will require the computation of $\langle Q_{8g} \rangle_0$ in
the large-$N$ approach or on the lattice.

\subsection{Missing low-energy contributions}

Two of the largest terms in our error budget concern low-energy physics:
hadronic matrix elements in the isospin limit, as well as corrections to the
isospin limit. In \cite{Pallante:1999qf,Pallante:2000hk} it has been pointed
out that for approaches that do not include final-state interactions, analyticity
suggests extra positive contributions to the value of $\bsi$ and negative
corrections to the value of $\bei$, both of which would raise $\epe$. (See
however \cite{Buras:2000kx}.)  If we naively apply the correction factors of
\cite{Pallante:1999qf,Pallante:2000hk} to typical large-$N$ values $\bsi=0.6$
and $\bei=0.8$, an increase of $\epe$ to $7.8 \times 10^{-4}$ results, still
well below the data. (Employing the lattice-inspired central values in our
error estimate, $\bsi=0.57$ and $\bei=0.76$, results in a very similar value
$\epe = 8.5 \times 10^{-4}$.) While a complete non-perturbative calculation
should account for the full matrix elements including final-state interactions,
the issue of final-state interactions may not yet be completely under
control\footnote{For instance, the final-state phase shifts obtained in
\cite{Bai:2015nea} are not in good agreement with the values extracted from
experiment. We thank Chris Sachrajda for discussion.} and certainly deserves
further study.

Another type of long-distance corrections is isospin breaking, both due to
electromagnetism and $m_u \ne m_d$. This is parametrised by the two parameters
$\hat\Omega_{\rm eff}$ and $a$. The latter only affects the overall
normalisation and cannot bring the SM into agreement with data. Explaining the
measured $\epe$ due to the former would require a value of opposite sign and
an order of magnitude larger than the value obtained in
\cite{Cirigliano:2003nn,Cirigliano:2003gt}. Nevertheless, given the profound
implications of the $\epe$ anomaly, this issue deserves further scrutiny,
and also lattice-QCD studies should take these corrections into account.

\subsection{Missing higher-order corrections to the Wilson coefficients}

Higher-order corrections to the Wilson coefficients will also have an impact
on the theory prediction of $\epe$. While it seems highly unlikely that they
can bring the SM prediction into agreement with experiment, it is still
instructive to discuss them in slightly more detail. In our analyses we fixed
the renormalisation scale to $\mu = m_c$ in the three-flavour theory.
Hence the computation of the  Wilson coefficients involves several steps, which
start with matching at the weak scale and end with integrating out the charm
quark at $\mu = m_c$. The intermediate steps involve the renormalisation group
evolution of $Q_1$ -- $Q_{10}$ and integrating out the bottom quark.  

The weak-scale matching corrections are known at NNLO for the electroweak
penguin \cite{Buras:1999st} as well as the current-current and QCD penguins
\cite{Bobeth:1999mk}, albeit in a different renormalisation scheme for the
later two. The respective scheme transformation is given in
\cite{Gorbahn:2004my}, where the relevant anomalous dimensions for the NNLO
evolution of $Q_1$ -- $Q_6$ can also be found. For these operators the matching
corrections at $\mu = m_b$ are also known \cite{Brod:2010mj}, yet all other
matching corrections and anomalous dimension matrices are currently known only
at NLO.

In particular the unknown matching corrections at $\mu=m_c$ could be sizeable
\cite{Brod:2011ty} since the strong coupling is growing rapidly in this region.
For this reason we estimated higher-order corrections by varying the matching
scale around $\mu= m_c$, and used the three-flavour renormalisation
group running to determine the Wilson coefficients at $\mu=m_c$. The resulting
residual scale dependence is typically in the ball park of 10\% for
$y_3$--$y_{10}$, but substantially smaller for $z_+$ and $z_-$. Using this
procedure, only the uncertainties in $y_6$, and to a lesser extent $y_8$, have
a significant impact on the error budget of $\epsilon'/\epsilon$.

The partially known NNLO corrections to $y_8$ are quite large
\cite{Buras:1999st} and decrease the SM prediction for $\epe$. Accordingly,
only the NNLO corrections to $y_6$ could arguably lead to a significant
enhancement of $\epsilon'/\epsilon$, but our error estimate shows that a 10\%
increase in $y_6$ results only in a $1.2 \times 10^{-4}$ increase in the SM
prediction. Bringing the SM prediction close to the experimental value would
require a very large higher-order correction to $y_6$ which would cast serious
doubts on the convergence of the perturbative series in our approach. If this
was indeed the case, we would have to perform our analysis in a four-flavour
setup, i.e. above the charm scale, which would also require new calculations
of matrix elements on the lattice.


\section {BSM physics in \boldmath{$\epe$}}
\label{sec:explBSM}

Not having been able to identify a plausible way to reconcile our prediction
with the data (other than attributing it to a large statistical fluctuation
somewhere), we turn to a discussion of physics Beyond the Standard Model (BSM)
in $\epe$. We first note that (\ref{eprime}), reproduced here for convenience:
\be
\frac{\varepsilon'}{\varepsilon} = -\,\frac{\omega_+}{\sqrt{2}\,|\varepsilon_K|}
\left[\, \frac{{\IM} A_0}{{\RE}A_0}\,(1-\hat\Omega_{\rm eff}) -
\frac{1}{a}\,\frac{{\IM} A_2}{{\RE}A_2} \,\right],
\ee
remains intact in the presence of new physics, which can be classified by
which of ${\rm Re} A_{0,2}$ and ${\rm Im} A_{0,2}$ is affected.

\begin{boldmath}
\subsection{BSM physics in ${\rm Re} A_{0,2}$}
\end{boldmath}

Noting that the RBC-UKQCD prediction  \cite{Bai:2015nea} 
of ${\rm Re} A_0$ exceeds the experimental
determination, while the large-$N$ method exhibits a deficit
\cite{Buras:2014maa},
we define the ratio 
\be
H=\frac{(\RE A_0)_\text{SM}}{(\RE A_0)_\text{EXP}},
\ee
which takes the central value $H=1.4$ and $H=0.7$ in  \cite{Bai:2015nea} and 
\cite{Buras:2014maa}, respectively. In other words, we are considering a
scenario where the experimental value of ${\rm Re} A_0$ is a sum of the SM
contribution and a BSM contribution. We cannot presently exclude that such a
sub-leading part of  $\RE A_0$ comes from NP, a possibility investigated in
\cite{Buras:2014sba}. As we have seen there is a strong correlation between
$\RE A_0$ and the matrix element of $Q_4$ and consequently there is an effect
on $\epe$. We stress that the denominators in (\ref{eprime}) are always the
true (experimental) values including any BSM contributions. It is the
\textit{numerator} term ${\rm Im} A_0$ that is affected through the
correlation of hadronic matrix elements.

Our formalism can easily be adapted to this case; one merely needs to
multiply the $V-A$ term given in  (\ref{ratioa}) by a factor of
$H$. In this fashion, the denominators in the ratios (\ref{ratioa})
are corrected for their BSM ``contamination'' and the theoretical
SM expressions are again valid. Note that the ratio (\ref{ratio6})
is not modified.
\begin{figure}[!t]
\begin{center}
\includegraphics[height=8cm]{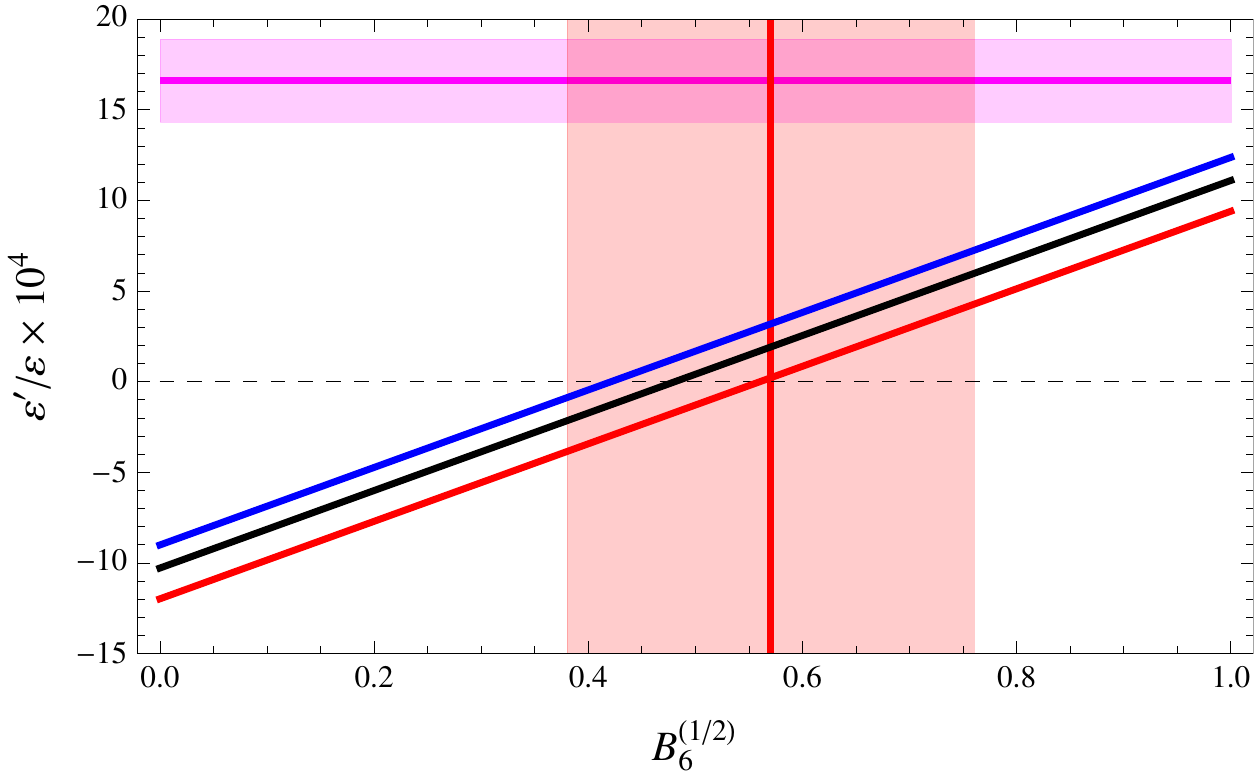}
\caption{$\epe$ as a function of $\bsi$, for three values of $H$
  defined in the text.
\label{fig:epeHB6} }
\end{center}
\end{figure}
In Figure~\ref{fig:epeHB6}, we plot $\epe$ as a function of $\bsi$ 
for $H=0.7$ (blue), $H=1.0$ (black), and $H=1.4$ (red). 
We see that taking the RBC-UKQCD central value for ${\rm Re} A_0$ to
be the true SM prediction, the agreement between theory and data for
$\epe$ is worsened -- and compensating for this requires even larger
values of $\bsi$ than in the SM. Conversely, taking the large-$N$
central value at face value one observes a slight improvement
(reduction) of the tension in $\epe$ by means of an upward shift.
But in both cases, the effect is not huge, dwarfed by the uncertainty
in $\bsi$, and reconciling theory and experiment still requires $\bsi>1$.
We conclude that CP-conserving data does not favour a scenario
of BSM in ${\rm Re} A_0$, although there is sizable room for it.
A similar discussion could be given for NP in ${\rm Re} A_2$.

\begin{boldmath}
\subsection{BSM physics in ${\rm Im} A_{0,2}$}
\end{boldmath}

The result obtained in our paper that $\epe$ in the SM is significantly below
the experimental data has an impact on various NP models. This is in particular
the case for models in which there is a strong correlation between $\epe$
and the branching ratios for rare decays $\kpn$ and $\klpn$. Such a correlation
has been stressed first in \cite{Buras:1998ed} and investigated in many papers
since then. See \cite{Buras:2013ooa} and references to earlier literature
therein.

In several models, like littlest Higgs model with T-parity (LHT)
\cite{Blanke:2009am}, and {generally  $Z$--models with new FCNCs,}
only in left-handed currents \cite{Buras:2012jb,Buras:2014sba}, enhancement of
the branching ratio for $\klpn$ is significantly constrained by $\epe$ because
in these models such an enhancement is correlated with the suppression of
$\epe$ with respect to the SM. This is also the case of $\kpn$ but as $\kpn$
receives in addition to imaginary parts of the relevant amplitudes also the
real parts, this correlation is much less pronounced. Therefore in such models
in order to have large enhancements of $\klpn$ and $\kpn$ the SM prediction for
$\epe$ must be above the data, which is certainly not favoured by our analysis.

Therefore, in these models the agreement with the data for $\epe$ can generally
be obtained only with strongly suppressed branching ratio for $\klpn$. In the
case of $\kpn$ this suppression is not required but significant departures from
the SM are not allowed. The recent analysis within the LHT model \cite{LHT15}
shows this explicitly.

Now, in the models just described NP enters $\epe$ only through ${\rm Im} A_{2}$
and the presence of only left-handed FCNCs implies uniquely the strict
correlation between $\epe$ and $\klpn$ mentioned above. But as shown in
\cite{Buras:2015yca} in the presence of both left-handed and right-handed FCNCs
it is possible to arrange these couplings without significant fine-tuning so
that the enhancement of $\epe$ required to fit data implies automatically the
enhancement of $\klpn$ and to lesser extent of $\kpn$. An explicit example of
a model with tree-level $Z$  exchanges contributing to $\epe$, $\klpn$ and
$\kpn$ can be found in \cite{Buras:2015yca}.

{In $Z^\prime$ models the situation can in principle be different even if
$\epe$ is only modified through ${\rm Im} A_{2}$ because flavour diagonal 
quark couplings to $Z^\prime$ could have proper signs so that $\epe$ and 
$\klpn$ can be simultaneously enhanced in models with only left-handed 
flavour violating $Z^\prime$ couplings. As pointed out in  
\cite{Buras:2014yna} some 331 models have this property.}

Another route towards the enhancement of $\epe$, less studied in the literature,
are $Z^\prime$ tree level exchanges with flavour universal diagonal couplings 
to quarks. In this case ${\rm Im} A_{2}$ is not modified and NP enters only 
${\rm Im} A_{0}$ through QCD penguin contributions. As demonstrated in 
\cite{Buras:2014sba,Buras:2015yca} also in this model $\epe$ and
$K\to\pi\nu\bar\nu$ can be simultaneously enhanced. Moreover, this can be
achieved with only left-handed FCNCs. If the $\epe$ anomaly will be confirmed
and future data on rare decays will exhibit such enhancements, models of this
kind and the ones mentioned in previous paragraph will be favoured.

Clearly there are other possibilities involving new operators, like
supersymmetric models \cite{Masiero:1999ub,Buras:1999da,Buras:2000qz},
Randall-Sundrum models \cite{Gedalia:2009ws}, or left-right models
\cite{Bertolini:2012pu}, but this is another story which requires further
study.


\section{Summary and outlook}\label{sec:5}

Motivated by the recent results on $K\to\pi\pi$ amplitudes from the RBC-UKQCD 
collaboration, we gave another look to the ratio $\epe$ within the SM. The
main result of our analysis is the identification of a possible new anomaly in
flavour physics, this time in $K$ physics. This was possible because:
\begin{itemize}
\item
Improved results for the parameters $\bsi$ and $\bei$ became available through
recent lattice-QCD studies by the RBC-UKQCD collaboration that are supported
by the large-$N$ approach which provides upper bounds on these parameters.
\item
We employed a formalism that is manifestly independent of the values
of leading $(V-A)\otimes(V-A)$ QCD penguin, and EW penguin hadronic matrix
elements of the operators $Q_4$, $Q_9$, and $Q_{10}$. In this manner a
prediction for $\epe$ could be made that is more precise than presently
possible by direct lattice-QCD simulations.
\end{itemize}

In this context, we have presented a new analytic formula for $\epe$ in terms
of $\bsi$ and $\bei$ that is valid also in SM extensions with the same
operator structure. This formula depends on the Wilson coefficients of the
contributing operators which are model dependent while $\bsi$ and $\bei$,
related to long-distance dynamics, are independent of NP contributions. Thus
our formula can be used for models such as the models with constrained MFV,
3-3-1 models and littlest Higgs models. We have also provided an update of a
formula for $\epe$ in which NP enters directly through the shifts in basic
one-loop functions.

Our analysis emphasises the correlation between the amplitude $\RE A_0$ and
the contribution of the $Q_4$ operator to $\epe$ given in (\ref{Q4}). As the
central value of $\RE A_0$ in \cite{Bai:2015nea} is by $40\%$ above the data,
this calculation overestimates the contribution of $Q_4$ to $\epe$ making it 
smaller. Assuming that $\RE A_0$ is fully described by SM dynamics, we
could improve the accuracy of its estimate by roughly an order of magnitude,
as seen in (\ref{Q4BJL}) and (\ref{Q4LATTICE}). 

We have extracted from \cite{Bai:2015nea} the value of $\bsi$ obtained by the
RBC-UKQCD collaboration to find that it is significantly lower than unity. In
fact this is the main reason for the low value of $\epe$ found in that paper.
On the other hand, should the values of $\bsi$ and $\bei$ eventually turn out
to be close to the upper bound from the large-$N$ approach \cite{Buras:2015xba},
significantly larger values of $\epe$ are found, although still roughly by
a factor of two below the data.

Our improved anatomy of $\epe$ clearly demonstrates that the SM has potential 
difficulties in describing the data for $\epe$. However, there are several open
questions that have to be answered before one can be fully confident that NP is
at work here. Answering them would also allow us to give a better estimate of
the room left for particular NP models.

Our analysis shows that the next most important issues that have to be 
clarified are as follows:
\begin{itemize}
\item
The value of $\bsi$ should be determined with an accuracy of at least $10\%$.
Fig.~\ref{fig:epeb6b8} demonstrates this need clearly, but also a higher
precision on $\bei$ would be beneficial.
\item
The values of the Wilson coefficients $y_i$ at the NNLO level. First steps in
this direction have been taken in \cite{Buras:1999st,Gorbahn:2004my}.
\item
Improved calculations of isospin breaking effects, represented in our
formula in (\ref{eprime}) by the parameters $a$ and $\hat\Omega_\text{eff}$. 
\item
The role of electromagnetic corrections to the hadronic matrix elements, as
emphasised already in \cite{Buras:1993dy}. Without these corrections there
remains some uncertainty due to the renormalisation scheme used for operators.
\item
Precise theoretical predictions of $\RE A_0$ and $\RE A_2$ within the SM, which
would tell us to which degree our assumption of neglecting NP contributions
in these amplitudes is justified.
\item
Finally, our understanding of the role of final-state interactions in $\epe$,
see \cite{Cirigliano:2011ny} and references therein, should be improved.
\end{itemize}

Our present results could be affected to some extent by the future finding that
some part of the amplitude $\RE A_0$ does not come from the SM dynamics but NP.
As Fig.~\ref{fig:epeHB6} shows, if $\RE A_0$ in the SM is below the experimental
value, as indicated by the large-$N$ approach \cite{Buras:2014maa}, the
suppression of $\epe$ by the $Q_4$ operator is smaller, implying a larger value
of $\epe$. On the other hand if $\RE A_0$ in the SM is above the experimental
value as presently seen in lattice data, the role of $Q_4$ will be enhanced
and consequently $\epe$ smaller. But as seen in the error budget of
Table~\ref{tab:errbud}, this effect is by far less important than the
sensitivity to $\bsi$.

In view of the tendency of $\epe$ in the SM to be significantly below the data,
it is exciting that in the coming years LHC might tell us what this physics
could be. But also independent studies of $\epe$ in various extensions of the
SM could select those extensions of the SM in which $\epe$ could be enhanced
over its SM value. In fact first phenomenological implications of our results
on new physics models have been presented in
\cite{Blanke:2015wba,Buras:2015yca}. In any case it appears that $\epe$ could
soon become again a leading light in flavour physics.

\section*{Acknowledgements}
We thank, Jean-Marc~G{\'e}rard, Chris~Kelly, Chris~Sachrajda and Amarjit~Soni
for discussions. Particular thanks go to Gerhard~Buchalla, Gino~Isidori,
Ulrich~Nierste and Jure~Zupan for inviting three of us to the MIAPP workshop
``Flavour 2015: New Physics at High Energy and High Precision'', during which
most of our analysis has been performed. S.J.\ also thanks the organisers of
the MIAPP workshop ``Anticipating 14 TeV: Insights into matter from the LHC
and beyond'' and the Excellence Cluster ``Universe'' for hospitality and a
stimulating work environment.
The research of AJB was done and financed in the context of the ERC Advanced
Grant project ``FLAVOUR''(267104) and was partially supported by the DFG
cluster of excellence ``Origin and Structure of the Universe''.
MG and SJ acknowledge support by the UK Science \& Technology Facilities 
Council (STFC) under grant numbers ST/L000431/1 and  ST/L000504/1
(respectively). SJ acknowledges the NExT institute.
MJ is supported in part by the Spanish Consolider-Ingenio 2010 Programme CPAN
(Grant number CSD2007-00042), by MINECO Grant numbers CICYT-FEDER-FPA2011-25948
and CICYT-FEDER-FPA2014-55613, by the Severo Ochoa excellence program of MINECO
under Grant number SO-2012-0234 and by Secretaria d'Universitats i Recerca del
Departament d'Economia i Coneixement de la Generalitat de Catalunya under
Grant number 2014 SGR 1450.

\appendix

\section{Subleading contributions to \boldmath{${\rm Im} A_{0,2}$} and related
 operator matrix elements.}\label{app:subleading}

Including operators with small Wilson coefficients and colour-suppressed
hadronic matrix element, the isospin ratios (\ref{ratioa}) and
(\ref{ratio6}) receive the following corrections:
\begin{eqnarray}
\Delta \left( \frac{ {\rm Im} A_0}{ {\rm Re} A_0} \right)_{V-A} 
&=&  {\rm Im} \tau \, \frac{p_3 \big( 2(y_3+y_4) - b\,(y_9+y_{10})\big)}{2 (1+q) z_-} \,,
\\
\Delta \left( \frac{ {\rm Im} A_0}{ {\rm Re} A_0} \right)_{6} \quad \;
&=&  -\,\frac{G_F}{\sqrt{2}}\, {\rm Im} \, \lambda_t \,
                \frac{ p_5\, y_5\, \langle Q_6 \rangle_0
              + b\,(y_8 + p_{70}\, y_7)\,\langle Q_8 \rangle_0 }{{\rm Re} A_0} \,,
\\
\end{eqnarray}
where we have defined
\begin{equation}
  p_3 =  \frac{ \langle Q_3 \rangle_0 } {\langle Q_- \rangle_0} \,,
  \qquad
  p_5 =  \frac{ \langle Q_5 \rangle_0 } {\langle Q_6 \rangle_0} \,,
 \qquad
  p_{70} = \frac{ \langle Q_7 \rangle_0 } {\langle Q_8 \rangle_0} \,.
\end{equation}
All three ratios are formally at least $1/N$-suppressed and multiplied by
small Wilson coefficients. Note that in (\ref{ratio8}) we have already included
the $y_7 \langle Q_7 \rangle_2$ contribution into $y_8^{\rm eff}$;
eq.~(\ref{ratio8}) then does not receive an additional correction.
We also define $B_8^{(1/2)}$ through
\be\label{eq:Q80}
\langle Q_8(\mu) \rangle_0= 2 h
\left[ \frac{m_{\rm K}^2}{m_s(\mu) + m_d(\mu)}\right]^2 F_\pi\,B_8^{(1/2)} \,.
\ee
This convention deviates from (5.12) in \cite{Buras:1993dy} but is
motivated by the result \cite{Buras:2015xba}
\be
B_8^{(1/2)}=1 + \ord \left( \frac{1}{N} \right)=1.1\pm0.1\,,
\ee
strongly supported by the lattice result for
$\langle Q_8(\mu) \rangle_0$ in \cite{Bai:2015nea}. 

To keep our phenomenological formulae simple and central values transparent,
we set the central values of $p_3$, $p_5$, and $p_{70}$ to zero and allow
generous error ranges that comprise both the intervals expected from large-$N$
counting and those computed in \cite{Bai:2015nea}. For the ratio $p_{72}$
defined below (\ref{y8eff}), which is also $1/N$-suppressed and plays a very
minor role numerically, we take the central value from \cite{Bai:2015nea} and,
conservatively, attribute a 100\% error to it. We furthermore employ
$B_8^{(1/2)}=1.0\pm 0.2 $, also derived from \cite{Bai:2015nea}.
All input rages are summarised in Table \ref{tab:inputs}. In this treatment,
we tend to overestimate our error on $\epe$, but as shown in the body of the
paper, this has a very minor impact on our predictions.

\section{Analytic Formula for \boldmath{$\epe$}}\label{app:X}

The expression (\ref{eq:8.3}) can be put into a formula that is more useful for
numerical evaluations as it shows explicitly the dependence on $m_t$ and $m_s$.
The most recent version of it has been presented in \cite{Buras:2014sba}, but
we update it here due to the change of some input parameters entering the
formulae for hadronic matrix elements  and a different treatment of isospin 
breaking corrections.
 We then have 
\be 
\left(\frac{\varepsilon'}{\varepsilon}\right)_{\rm SM}= \,\IM\lambda_t
\cdot F_{\varepsilon'}(x_t) \,,
\label{epeth}
\ee
where 
\be
F_{\varepsilon'}(x_t) =P_0 + P_X \, X_0(x_t) + 
P_Y \, Y_0(x_t) + P_Z \, Z_0(x_t)+ P_E \, E_0(x_t)~.
\label{FE}
\ee
The first term is dominated by QCD-penguin contributions, the next three
terms by electroweak penguin contributions. The last term expresses the
$m_t$ dependence from contribution of QCD penguin operators and is totally
negligible. The $x_t$ dependent functions are given as follows
\begin{equation}\label{X01}
X_0(x_t) = \frac{x_t}{8}\left[\frac{x_t+2}{x_t-1} + \frac{3x_t-6}{(x_t-1)^2}
\log x_t\right] ,
\end{equation}
\begin{equation}\label{Y0}
Y_0(x_t)={\frac{x_t}{8}}\left[{\frac{x_t -4}{x_t-1}} 
+ {\frac{3 x_t}{(x_t -1)^2}} \ln x_t\right] ,
\end{equation}
\bea\label{Z0}
Z_0(x_t)~\!\!\!\!&=&\!\!\!\!~-\,{\frac{1}{9}}\ln x_t + 
{\frac{18x_t^4-163x_t^3 + 259x_t^2-108x_t}{144 (x_t-1)^3}}+
\nonumber\\ 
&&\!\!\!\!+\,{\frac{32x_t^4-38x_t^3-15x_t^2+18x_t}{72(x_t-1)^4}}\ln x_t \,,
\eea
\begin{equation}\label{E0}
E_0(x_t)=-\,{\frac{2}{3}}\ln x_t+{\frac{x_t^2(15-16x_t+4x_t^2)}{6(1-x_t)^4}}
\ln x_t+{\frac{x_t(18-11x_t-x_t^2)}{12(1-x_t)^3}} ~,
\end{equation}
where $x_t=m^2_t/M_W^2$.

The coefficients $P_i$ are given by
\begin{equation}
P_i = r_i^{(0)} + 
r_i^{(6)} R_6 + r_i^{(8)} R_8 \,,
\label{eq:pbePi}
\end{equation}
where we have defined
\begin{align}\label{RS}
R_6&\equiv \bsi\left[ \frac{114.54\mev}{m_s(m_c)+m_d(m_c)} \right]^2,\\
R_8&\equiv \bei\left[ \frac{114.54\mev}{m_s(m_c)+m_d(m_c)} \right]^2.
\end{align}

The coefficients $r_i^{(0)}$, $r_i^{(6)}$ and $r_i^{(8)}$ comprise information
on the Wilson-coefficient functions of the $\Delta S=1$ weak effective
Hamiltonian at NLO and incorporate the values of the matrix elements of those
operators that we could extract by imposing the experimental values of
${\rm Re} A_0$ and ${\rm Re} A_2$. Their numerical values are given in the
NDR-${\rm \overline{MS}}$ renormalisation scheme for $\mu=m_c$ and three
values of $\alpha_s(M_Z)$ in Table~\ref{tab:pbendr}.

\begin{table}[!tb]
\begin{center}
\begin{tabular}{|c||c|c|c||c|c|c||c|c|c|}
\hline
&\multicolumn{3}{c||}{$\alpha_s(M_Z)= 0.1179$}&
 \multicolumn{3}{c||}{$\alpha_s(M_Z)= 0.1185$} &
  \multicolumn{3}{c|}{$\alpha_s(M_Z)= 0.1191$} \\
  \hline
$i$ & $r_i^{(0)}$ & $r_i^{(6)}$ & $r_i^{(8)}$ &
      $r_i^{(0)}$ & $r_i^{(6)}$ & $r_i^{(8)}$ &
       $r_i^{(0)}$ & $r_i^{(6)}$ & $r_i^{(8)}$ \\
      \hline
0 &    -3.392 & 15.293 & 1.271 &
       -3.421 & 15.624 & 1.231 &
       -3.451 & 15.967 & 1.191 \\
$X_0$ & 0.655 &  0.029 & 0. &
        0.655 &  0.030 & 0. &
        0.655 &  0.031 & 0. \\
$Y_0$ & 0.451 &  0.114 & 0. &
        0.449 &  0.116 & 0. &
        0.447 &  0.118 & 0. \\
$Z_0$ & 0.406 & -0.022 & -13.434 &
        0.420 & -0.022 & -13.649 &
        0.435 & -0.023 & -13.872 \\
$E_0$ & 0.229 & -1.760 & 0.652 &
        0.228 & -1.788 & 0.665 &
        0.226 & -1.816 & 0.678 \\
\hline
\end{tabular}
\end{center}
\caption{\it The coefficients $r_i^{(0)}$, $r_i^{(6)}$ and $r_i^{(8)}$ of
formula (\ref{eq:pbePi}) in the  NDR-${\rm \overline{MS}}$  scheme for three values of $\alpha_s(M_Z)$.
\label{tab:pbendr}}~\\[-2mm]\hrule
\end{table}

\renewcommand{\refname}{R\lowercase{eferences}}

\addcontentsline{toc}{section}{References}


\begin{thebibliography}{10}

\bibitem{Bertolini:1998vd}
S.~Bertolini, M.~Fabbrichesi, and J.~O. Eeg, {\it {Theory of the CP violating
  parameter $\epsilon'/\epsilon$}},  {\em Rev. Mod. Phys.} {\bf 72} (2000)
  65--93, [\href{http://arxiv.org/abs/hep-ph/9802405}{{\tt hep-ph/9802405}}].

\bibitem{Buras:2003zz}
A.~J. Buras and M.~Jamin, {\it {$\varepsilon'/\varepsilon$ at the NLO: 10 years
  later}},  {\em JHEP} {\bf 01} (2004) 048,
  [\href{http://arxiv.org/abs/hep-ph/0306217}{{\tt hep-ph/0306217}}].

\bibitem{Pich:2004ee}
A.~Pich, {\it {$\varepsilon'/\varepsilon$ in the Standard Model: Theoretical
  update}},  \href{http://arxiv.org/abs/hep-ph/0410215}{{\tt hep-ph/0410215}}.

\bibitem{Cirigliano:2011ny}
V.~Cirigliano, G.~Ecker, H.~Neufeld, A.~Pich, and J.~Portoles, {\it {Kaon
  Decays in the Standard Model}},  {\em Rev. Mod. Phys.} {\bf 84} (2012) 399,
  [\href{http://arxiv.org/abs/1107.6001}{{\tt arXiv:1107.6001}}].

\bibitem{Bertolini:2012pu}
S.~Bertolini, J.~O. Eeg, A.~Maiezza, and F.~Nesti, {\it {New physics in
  $\varepsilon'$ from gluomagnetic contributions and limits on Left-Right
  symmetry}},  {\em Phys. Rev.} {\bf D86} (2012) 095013,
  [\href{http://arxiv.org/abs/1206.0668}{{\tt arXiv:1206.0668}}].

\bibitem{Buras:1991jm}
A.~J. Buras, M.~Jamin, M.~Lautenbacher, and P.~H. Weisz, {\it {Effective
  Hamiltonians for $\Delta S = 1$ and $\Delta B = 1$ non-leptonic decays beyond
  the leading logarithmic approximation}},  {\em Nucl. Phys.} {\bf B370} (1992)
  69--104.

\bibitem{Buras:1992tc}
A.~J. Buras, M.~Jamin, M.~E. Lautenbacher, and P.~H. Weisz, {\it {Two loop
  anomalous dimension matrix for $\Delta S = 1$ weak non-leptonic decays. 1.
  $\ord(\alpha_s^2)$}},  {\em Nucl. Phys.} {\bf B400} (1993) 37--74,
  [\href{http://arxiv.org/abs/hep-ph/9211304}{{\tt hep-ph/9211304}}].

\bibitem{Buras:1992zv}
A.~J. Buras, M.~Jamin, and M.~E. Lautenbacher, {\it {Two-loop anomalous
  dimension matrix for $\Delta S = 1$ weak non-leptonic decays. 2.
  $\ord(\alpha\alpha_s)$}},  {\em Nucl. Phys.} {\bf B400} (1993) 75--102,
  [\href{http://arxiv.org/abs/hep-ph/9211321}{{\tt hep-ph/9211321}}].

\bibitem{Ciuchini:1992tj}
M.~Ciuchini, E.~Franco, G.~Martinelli, and L.~Reina, {\it
  {$\varepsilon'/\varepsilon$ at the next-to-leading order in QCD and QED}},
  {\em Phys. Lett.} {\bf B301} (1993) 263--271,
  [\href{http://arxiv.org/abs/hep-ph/9212203}{{\tt hep-ph/9212203}}].

\bibitem{Buras:1993dy}
A.~J. Buras, M.~Jamin, and M.~E. Lautenbacher, {\it The anatomy of
  $\varepsilon'/ \varepsilon$ beyond leading logarithms with improved hadronic
  matrix elements},  {\em Nucl. Phys.} {\bf B408} (1993) 209--285,
  [\href{http://arxiv.org/abs/hep-ph/9303284}{{\tt hep-ph/9303284}}].

\bibitem{Ciuchini:1993vr}
M.~Ciuchini, E.~Franco, G.~Martinelli, and L.~Reina, {\it {The $\Delta S = 1$
  effective Hamiltonian including next-to-leading order QCD and QED
  corrections}},  {\em Nucl. Phys.} {\bf B415} (1994) 403--462,
  [\href{http://arxiv.org/abs/hep-ph/9304257}{{\tt hep-ph/9304257}}].

\bibitem{Buras:1999st}
A.~J. Buras, P.~Gambino, and U.~A. Haisch, {\it {Electroweak penguin
  contributions to non-leptonic $\Delta F=1$ decays at NNLO}},  {\em Nucl.
  Phys.} {\bf B570} (2000) 117--154,
  [\href{http://arxiv.org/abs/hep-ph/9911250}{{\tt hep-ph/9911250}}].

\bibitem{Gorbahn:2004my}
M.~Gorbahn and U.~Haisch, {\it {Effective Hamiltonian for non-leptonic $|\Delta
  F| = 1$ decays at NNLO in QCD}},  {\em Nucl. Phys.} {\bf B713} (2005)
  291--332, [\href{http://arxiv.org/abs/hep-ph/0411071}{{\tt hep-ph/0411071}}].

\bibitem{Brod:2010mj}
J.~Brod and M.~Gorbahn, {\it {$\epsilon_K$ at Next-to-Next-to-Leading Order:
  The Charm-Top-Quark Contribution}},  {\em Phys. Rev.} {\bf D82} (2010)
  094026, [\href{http://arxiv.org/abs/1007.0684}{{\tt arXiv:1007.0684}}].

\bibitem{Buchalla:1989we}
G.~Buchalla, A.~J. Buras, and M.~K. Harlander, {\it The anatomy of
  $\varepsilon' / \varepsilon$ in the standard model},  {\em Nucl. Phys.} {\bf
  B337} (1990) 313--362.

\bibitem{Bardeen:1986uz}
W.~A. Bardeen, A.~J. Buras, and J.-M. G\'erard, {\it {The $K\to\pi \pi$ Decays
  in the Large-N Limit: Quark Evolution}},  {\em Nucl. Phys.} {\bf B293} (1987)
  787.

\bibitem{Buras:2014maa}
A.~J. Buras, J.-M. G{\'e}rard, and W.~A. Bardeen, {\it {Large-$N$ Approach to
  Kaon Decays and Mixing 28 Years Later: $\Delta I = 1/2$ Rule, $\hat B_K$ and
  $\Delta M_K$}},  {\em Eur. Phys. J.} {\bf C74} (2014), no.~5 2871,
  [\href{http://arxiv.org/abs/1401.1385}{{\tt arXiv:1401.1385}}].

\bibitem{Buras:1996dq}
A.~J. Buras, M.~Jamin, and M.~E. Lautenbacher, {\it {A 1996 analysis of the CP
  violating ratio $\varepsilon'/\varepsilon$}},  {\em Phys. Lett.} {\bf B389}
  (1996) 749--756, [\href{http://arxiv.org/abs/hep-ph/9608365}{{\tt
  hep-ph/9608365}}].

\bibitem{Bosch:1999wr}
S.~Bosch et~al., {\it Standard model confronting new results for
  $\varepsilon'/\varepsilon$},  {\em Nucl. Phys.} {\bf B565} (2000) 3--37,
  [\href{http://arxiv.org/abs/hep-ph/9904408}{{\tt hep-ph/9904408}}].

\bibitem{Buras:2014sba}
A.~J. Buras, F.~De~Fazio, and J.~Girrbach, {\it {$\Delta I=1/2$ rule,
  $\varepsilon '/\varepsilon $ and $K\rightarrow \pi \nu \bar{\nu }$ in $Z'
  (Z)$ and $G' $ models with FCNC quark couplings}},  {\em Eur. Phys. J.} {\bf
  C74} (2014) 2950, [\href{http://arxiv.org/abs/1404.3824}{{\tt
  arXiv:1404.3824}}].

\bibitem{Buras:2015qea}
A.~J. Buras, D.~Buttazzo, J.~Girrbach-Noe, and R.~Knegjens, {\it
  {$K^+\to\pi^+\nu\bar\nu$ and $K_L\to\pi^0\nu\bar\nu$ in the Standard Model:
  Status and Perspectives}},  \href{http://arxiv.org/abs/1503.02693}{{\tt
  arXiv:1503.02693}}.

\bibitem{Hambye:1998sma}
T.~Hambye, G.~Kohler, E.~Paschos, P.~Soldan, and W.~A. Bardeen, {\it {$1 / N$
  corrections to the hadronic matrix elements of $Q_6$ and $Q_8$ in $K\to\pi
  \pi$ decays}},  {\em Phys. Rev.} {\bf D58} (1998) 014017,
  [\href{http://arxiv.org/abs/hep-ph/9802300}{{\tt hep-ph/9802300}}].

\bibitem{Blum:2015ywa}
T.~Blum, P.~Boyle, N.~Christ, J.~Frison, N.~Garron, et~al., {\it {$K
  \rightarrow \pi\pi$ $\Delta I=3/2$ decay amplitude in the continuum limit}},
  \href{http://arxiv.org/abs/1502.00263}{{\tt arXiv:1502.00263}}.

\bibitem{Buras:2015xba}
A.~J. Buras and J.-M. G\'erard, {\it {Upper Bounds on
  $\varepsilon'/\varepsilon$ Parameters $B_6^{(1/2)}$ and $B_8^{(3/2)}$ from
  Large-$N$ approach and other News}},
  \href{http://arxiv.org/abs/1507.06326}{{\tt arXiv:1507.06326}}.

\bibitem{Bai:2015nea}
Z.~Bai, T.~Blum, P.~Boyle, N.~Christ, J.~Frison, et~al., {\it {Standard-model
  prediction for direct CP violation in $K\to\pi\pi$ decay}},
  \href{http://arxiv.org/abs/1505.07863}{{\tt arXiv:1505.07863}}.

\bibitem{Batley:2002gn}
{\bf NA48} Collaboration, J.~Batley et~al., {\it {A Precision measurement of
  direct CP violation in the decay of neutral kaons into two pions}},  {\em
  Phys. Lett.} {\bf B544} (2002) 97--112,
  [\href{http://arxiv.org/abs/hep-ex/0208009}{{\tt hep-ex/0208009}}].

\bibitem{AlaviHarati:2002ye}
{\bf KTeV} Collaboration, A.~Alavi-Harati et~al., {\it {Measurements of direct
  CP violation, CPT symmetry, and other parameters in the neutral kaon
  system}},  {\em Phys. Rev.} {\bf D67} (2003) 012005,
  [\href{http://arxiv.org/abs/hep-ex/0208007}{{\tt hep-ex/0208007}}].

\bibitem{Worcester:2009qt}
{\bf KTeV} Collaboration, E.~Worcester, {\it {The Final Measurement of
  $\varepsilon'/\varepsilon$ from KTeV}},
  \href{http://arxiv.org/abs/0909.2555}{{\tt arXiv:0909.2555}}.

\bibitem{Cirigliano:2003gt}
V.~Cirigliano, G.~Ecker, H.~Neufeld, and A.~Pich, {\it {Isospin breaking in
  $K\to\pi\pi$ decays}},  {\em Eur. Phys. J.} {\bf C33} (2004) 369--396,
  [\href{http://arxiv.org/abs/hep-ph/0310351}{{\tt hep-ph/0310351}}].

\bibitem{Cirigliano:2003nn}
V.~Cirigliano, A.~Pich, G.~Ecker, and H.~Neufeld, {\it {Isospin violation in
  $\epsilon^\prime$}},  {\em Phys. Rev. Lett.} {\bf 91} (2003) 162001,
  [\href{http://arxiv.org/abs/hep-ph/0307030}{{\tt hep-ph/0307030}}].

\bibitem{Buras:1985yx}
A.~J. Buras and J.-M. G\'erard, {\it {$1/N$ Expansion for Kaons}},  {\em
  Nucl.Phys.} {\bf B264} (1986) 371.

\bibitem{Buras:1987wc}
A.~J. Buras and J.~M. G\'erard, {\it {Isospin Breaking Contributions to
  $\epe$}},  {\em Phys. Lett.} {\bf B192} (1987) 156.

\bibitem{Agashe:2014kda}
{\bf Particle Data Group} Collaboration, K.~Olive et~al., {\it {Review of
  Particle Physics}},  {\em Chin.Phys.} {\bf C38} (2014) 090001. {Updates
  available on \texttt{http://pdg.lbl.gov}.}

\bibitem{Aoki:2013ldr}
S.~Aoki, Y.~Aoki, C.~Bernard, T.~Blum, G.~Colangelo, et~al., {\it {Review of
  lattice results concerning low-energy particle physics}},  {\em Eur. Phys.
  J.} {\bf C74} (2014), no.~9 2890, [\href{http://arxiv.org/abs/1310.8555}{{\tt
  arXiv:1310.8555}}].

\bibitem{Blum:2012uk}
T.~Blum, P.~Boyle, N.~Christ, N.~Garron, E.~Goode, et~al., {\it {Lattice
  determination of the $K \to (\pi\pi)_{I=2}$ Decay Amplitude $A_2$}},  {\em
  Phys. Rev.} {\bf D86} (2012) 074513,
  [\href{http://arxiv.org/abs/1206.5142}{{\tt arXiv:1206.5142}}].

\bibitem{UTfit}
{\bf UTfit} Collaboration. {http://www.utfit.org}.

\bibitem{Charles:2015gya}
J.~Charles, O.~Deschamps, S.~Descotes-Genon, H.~Lacker, A.~Menzel, et~al., {\it
  {Current status of the Standard Model CKM fit and constraints on $\Delta F=2$
  New Physics}},  \href{http://arxiv.org/abs/1501.05013}{{\tt
  arXiv:1501.05013}}. {Updates on
  \href{http://ckmfitter.in2p3.fr}{http://ckmfitter.in2p3.fr}}.

\bibitem{Brod:2011ty}
J.~Brod and M.~Gorbahn, {\it {Next-to-Next-to-Leading-Order Charm-Quark
  Contribution to the CP Violation Parameter $\varepsilon_K$ and $\Delta
  M_K$}},  {\em Phys.Rev.Lett.} {\bf 108} (2012) 121801,
  [\href{http://arxiv.org/abs/1108.2036}{{\tt arXiv:1108.2036}}].

\bibitem{Bertolini:1993rc}
S.~Bertolini, M.~Fabbrichesi, and E.~Gabrielli, {\it {The Relevance of the
  dipole Penguin operators in $\varepsilon'/\varepsilon$}},  {\em Phys. Lett.}
  {\bf B327} (1994) 136--144, [\href{http://arxiv.org/abs/hep-ph/9312266}{{\tt
  hep-ph/9312266}}].

\bibitem{Buras:1999da}
A.~J. Buras, G.~Colangelo, G.~Isidori, A.~Romanino, and L.~Silvestrini, {\it
  {Connections between $\varepsilon'/\varepsilon$ and rare kaon decays in
  supersymmetry}},  {\em Nucl. Phys.} {\bf B566} (2000) 3--32,
  [\href{http://arxiv.org/abs/hep-ph/9908371}{{\tt hep-ph/9908371}}].

\bibitem{Pallante:1999qf}
E.~Pallante and A.~Pich, {\it {Strong enhancement of $\varepsilon'/\varepsilon$
  through final state interactions}},  {\em Phys. Rev. Lett.} {\bf 84} (2000)
  2568--2571, [\href{http://arxiv.org/abs/hep-ph/9911233}{{\tt
  hep-ph/9911233}}].

\bibitem{Pallante:2000hk}
E.~Pallante and A.~Pich, {\it {Final state interactions in kaon decays}},  {\em
  Nucl. Phys.} {\bf B592} (2001) 294--320,
  [\href{http://arxiv.org/abs/hep-ph/0007208}{{\tt hep-ph/0007208}}].

\bibitem{Buras:2000kx}
A.~J. Buras et~al., {\it {Final state interactions and
  $\varepsilon'/\varepsilon$: A critical look}},  {\em Phys. Lett.} {\bf B480}
  (2000) 80--86, [\href{http://arxiv.org/abs/hep-ph/0002116}{{\tt
  hep-ph/0002116}}].

\bibitem{Bobeth:1999mk}
C.~Bobeth, M.~Misiak, and J.~Urban, {\it {Photonic penguins at two loops and
  $m_t$-dependence of $BR(B\to X_s \ell^+ \ell^-)$}},  {\em Nucl. Phys.} {\bf
  B574} (2000) 291--330, [\href{http://arxiv.org/abs/hep-ph/9910220}{{\tt
  hep-ph/9910220}}].

\bibitem{Buras:1998ed}
A.~J. Buras and L.~Silvestrini, {\it {Upper bounds on $K \to\pi\nu\bar\nu$ and
  $K_L\to\pi^0 e^+e^-$ from $\varepsilon'/\varepsilon$ and $K_L \to\mu^+
  \mu^-$}},  {\em Nucl. Phys.} {\bf B546} (1999) 299--314,
  [\href{http://arxiv.org/abs/hep-ph/9811471}{{\tt hep-ph/9811471}}].

\bibitem{Buras:2013ooa}
A.~J. Buras and J.~Girrbach, {\it {Towards the Identification of New Physics
  through Quark Flavour Violating Processes}},  {\em Rept. Prog. Phys.} {\bf
  77} (2014) 086201, [\href{http://arxiv.org/abs/1306.3775}{{\tt
  arXiv:1306.3775}}].

\bibitem{Blanke:2009am}
M.~Blanke, A.~J. Buras, B.~Duling, S.~Recksiegel, and C.~Tarantino, {\it {FCNC
  Processes in the Littlest Higgs Model with T-Parity: a 2009 Look}},  {\em
  Acta Phys.Polon.} {\bf B41} (2010) 657--683,
  [\href{http://arxiv.org/abs/0906.5454}{{\tt arXiv:0906.5454}}].

\bibitem{Buras:2012jb}
A.~J. Buras, F.~De~Fazio, and J.~Girrbach, {\it {The Anatomy of Z' and Z with
  Flavour Changing Neutral Currents in the Flavour Precision Era}},  {\em JHEP}
  {\bf 1302} (2013) 116, [\href{http://arxiv.org/abs/1211.1896}{{\tt
  arXiv:1211.1896}}].

\bibitem{LHT15}
M.~Blanke, A.~J. Buras, and S.~Recksiegel, {\it {Quark flavour observables in
  the Littlest Higgs model with T-parity after LHC Run 1}},
  \href{http://arxiv.org/abs/1507.06316}{{\tt arXiv:1507.06316}}.

\bibitem{Buras:2015yca}
A.~J. Buras, D.~Buttazzo, and R.~Knegjens, {\it {$K\to\pi\nu\bar\nu$ and $\epe$
  in Simplified New Physics Models}},
  \href{http://arxiv.org/abs/1507.08672}{{\tt arXiv:1507.08672}}.

\bibitem{Buras:2014yna}
A.~J. Buras, F.~De~Fazio, and J.~Girrbach-Noe, {\it {Z-Z' mixing and Z-mediated
  FCNCs in $SU(3)_C \times SU(3)_L \times U(1)_X$ Models}},  {\em JHEP} {\bf
  1408} (2014) 039, [\href{http://arxiv.org/abs/1405.3850}{{\tt
  arXiv:1405.3850}}].

\bibitem{Masiero:1999ub}
A.~Masiero and H.~Murayama, {\it {Can $\varepsilon'/\varepsilon$ be
  supersymmetric?}},  {\em Phys. Rev. Lett.} {\bf 83} (1999) 907--910,
  [\href{http://arxiv.org/abs/hep-ph/9903363}{{\tt hep-ph/9903363}}].

\bibitem{Buras:2000qz}
A.~J. Buras, P.~Gambino, M.~Gorbahn, S.~Jager, and L.~Silvestrini, {\it
  $\varepsilon'/\varepsilon$ and rare $k$ and $b$ decays in the mssm},  {\em
  Nucl. Phys.} {\bf B592} (2001) 55--91,
  [\href{http://arxiv.org/abs/hep-ph/0007313}{{\tt hep-ph/0007313}}].

\bibitem{Gedalia:2009ws}
O.~Gedalia, G.~Isidori, and G.~Perez, {\it {Combining Direct and Indirect Kaon
  CP Violation to Constrain the Warped KK Scale}},  {\em Phys.Lett.} {\bf B682}
  (2009) 200--206, [\href{http://arxiv.org/abs/0905.3264}{{\tt
  arXiv:0905.3264}}].

\bibitem{Blanke:2015wba}
M.~Blanke, A.~J. Buras, and S.~Recksiegel, {\it {Quark flavour observables in
  the Littlest Higgs model with T-parity after LHC Run 1}},
  \href{http://arxiv.org/abs/1507.06316}{{\tt arXiv:1507.06316}}.

\end{thebibliography}
\providecommand{\href}[2]{#2}\begingroup\raggedright\endgroup
\end{document}